\documentclass[letterpaper,english,reprint,nofootinbib,aps,superscriptaddress,showpacs,showkeys]{revtex4-1}

\usepackage{babel,calc,amsmath,amsthm,amssymb,graphicx,subfigure,xcolor,comment}
\usepackage{mathdots}
\usepackage[T1]{fontenc}
\usepackage[unicode=true]{hyperref}
\hypersetup{
	colorlinks=true,       		% false: boxed links; true: colored links
	linkcolor=red,          	% color of internal links
	citecolor=blue,             % color of links to bibliography
	urlcolor=blue,          % color of external links
}

\usepackage{booktabs}

\usepackage{makecell}
\usepackage{float}
\usepackage{caption}
\usepackage{braket}
\usepackage[normalem]{ulem}

\def\C{\mathbb{C}}

\def\ot{\otimes}

\def\H{\mathcal{H}}
\def\D{\mathcal{D}}
\def\<{\langle}
\def\>{\rangle}
\def\lm{\lambda}
\newif\ifdebug

\debugtrue

\ifdebug
\definecolor{zhliu}{rgb}{0.48, 0.12, 0}

\newcommand{\note}[1]{\textcolor{zhliu}{#1}}
\newcommand\delete{\bgroup\markoverwith{\textcolor{zhliu}{\rule[0.5ex]{2pt}{0.8pt}}}\ULon}

\else

\newcommand{\note}[1]{\ignorespaces}
\newcommand{\delete}[1]{\ignorespaces}

\fi

\begin{document}

	\renewcommand{\figurename}{Fig.}
	
    \title{Correlation-Enhanced Quantum Reservoir Processing for Two-Qutrit Steering: From Generalization to Discovery}
	\author{Pu Wang}
	\affiliation{
		School of Control and Computer Engineering,
		North China Electric Power University,
		Beijing 102206, China
	}
	
	\author{Zhongyan Li}
 \email{lzhongy@ncepu.edu.cn}
	\affiliation{
		School of Mathematics and Physics,
		North China Electric Power University,
		Beijing 102206, China
	}

		\author{Huixian Meng}
	\email{huixianmenghd@ncepu.edu.cn}
	\affiliation{
		School of Mathematics and Physics,
		North China Electric Power University,
		Beijing 102206, China}
\author{Yutao Xin}
	\affiliation{
		School of Mathematics and Physics,
		North China Electric Power University,
		Beijing 102206, China}
\author{Ziqi Zhi}
	\affiliation{
		School of Mathematics and Physics,
		North China Electric Power University,
		Beijing 102206, China}
\author{Zhihua Guo}
\email{guozhihua@snnu.edu.cn}
	\affiliation{
		School of Mathematics and Statistics, Shaanxi Normal University,
		Xi'an 710119, China}
\author{Huaixin Cao}
\email{caohx@snnu.edu.cn}
	\affiliation{
		School of Artificial Intelligence, Wenshan University,
		Wenshan 663000, China}
	
	\date{\today}

\begin{abstract}
Quantum reservoir processing (QRP) has been applied to entanglement quantification, yet its potential for generalizable steering detection and the discovery of high-dimensional quantum correlations remains largely unexplored. Here, we introduce correlation-enhanced QRP (CEQRP), extending QRP to two-qutrit systems with two-body spin-correlation readouts. Some CEQRP-based classifiers substantially improve cross-family generalization, with logistic regression trained on full-information CEQRP features achieving approximately $96\%$ accuracy on unseen Werner states and over $97\%$  accuracy on unseen isotropic states at both \(m=4\) and \(m=5\). Independently, we establish two interconnected steering criteria for mixed two-qutrit states: a necessary-and-sufficient all-versus-nothing criterion for states admitting three linearly independent pure conditional states, and an asymptotic sufficient criterion, with the two criteria linked through invertible local filtering. These criteria provide rigorously certified steerable samples for training-set augmentation. Combining CEQRP-based classification with these theory-certified samples enables the discovery of new Alice-to-Bob steerable partially entangled states, including states at \(p=0.45\) whose steerability is independently confirmed by semidefinite programming (SDP) using seven projective measurement settings. The trained classifiers further guide finite-setting searches for steering boundaries. Our results establish a pathway from generalizable classification to the discovery of high-dimensional quantum steering.
\end{abstract}
	
%	\pacs{03.65.Ud, 03.67.Mn, 42.50.Xa}
	
\maketitle

\section{Introduction}\label{section1}

Quantum steering is a distinct form of quantum
correlation that lies strictly between entanglement
and Bell nonlocality
\cite{RevModPhys.92.015001,Jones,Wiseman}. It describes how one party can remotely prepare conditional states for another by performing local measurements, when the measuring party's devices are untrusted. This directional form of quantum correlation is relevant to one-sided device-independent tasks, including quantum key distribution, randomness certification, and quantum
communication \cite{ref2,ref3,ref4,ref5,PhysRevLett.134.160803,
	ManZhongXiao2005,Tian}. Determining whether a state is steerable and identifying the measurement resources needed to reveal steering are therefore fundamental problems.
%It describes the ability of one party to remotely
%prepare conditional quantum states for another
%through local measurements, with the former
%party's measurement devices treated as untrusted.
%This directional character makes steering a
%valuable resource for one-sided device-independent
%quantum information tasks, including quantum key
%distribution, randomness certification, and
%quantum communication
%\cite{ref2,ref3,ref4,ref5,PhysRevLett.134.160803,
%	ManZhongXiao2005,Tian}.
%Determining the steerability of quantum states
%and the measurement resources required to
%reveal it are therefore fundamental problems.

For a bipartite state $\rho_{AB}$,
measurements $\{M_{a|x}\}$ performed by Alice
generate an assemblage of unnormalized
conditional states on Bob,
\begin{equation}
	\sigma_{a|x}
	=
	\operatorname{Tr}_A
	\left[
	(M_{a|x}\otimes I_B)\rho_{AB}
	\right].
	\notag
\end{equation}
For a given measurement set, steering is demonstrated when the resulting assemblage admits no local-hidden-state (LHS) model. Various methods have been developed for
steering detection, including steering
inequalities, uncertainty-relation criteria,
explicit LHS constructions, and
semidefinite programming (SDP)
\cite{ref13,ref14,ref15,PhysRevA.93.022121,
	ref20,PhysRevA.97.012130,Guo,luo}.
In particular, SDP provides a systematic
approach to certifying steering for finite
sets of measurements.
However, failure to detect steering
with a limited number of settings does not
establish unsteerability under arbitrary
measurements. Increasing the measurement
resources can improve detection but also
increases the computational cost.

These challenges become more pronounced
in high-dimensional systems.
While two-qubit steering has been extensively
studied \cite{Nguyen_2019}, the steering
properties of mixed two-qutrit states remain
incompletely characterized.
The larger state space and richer correlation structure of two-qutrit systems complicate both analytical characterization and numerical optimization.
Machine learning (ML) offers a complementary
approach by learning steering classifications
from labeled quantum states
\cite{ref23,ref24,ref25,ref26,Jia,ZhangQ,SU2026112134}.
Nevertheless, classifiers trained on one
collection of states may fail to generalize
to structurally distinct state families.
Reliable steering labels and transferable
feature representations are therefore
essential for ML-based steering detection.

Quantum reservoir processing (QRP) offers
a promising route to constructing dynamical
features for quantum learning
\cite{Ghosh2019,sciadvady7987,Innocenti,Krisnanda}.
In QRP, an input state is processed by a fixed quantum reservoir, and time-dependent observables of the reservoir are used as features for a classical learner.
QRP has been applied to entanglement
quantification in two-qubit systems
\cite{PhysRevA.111.022412}.
However, its potential for detecting
high-dimensional quantum steering, particularly
its ability to generalize across state
families and identify steerable states
beyond the scope of existing criteria,
remains largely unexplored.

In this work, we investigate two-qutrit steering by combining correlation-enhanced quantum reservoir features with theory-certified training samples, aiming to move beyond generalizable classification toward identifying new steerable states. Our approach builds on two complementary advances.

First, we introduce correlation-enhanced quantum reservoir processing (CEQRP), a feature-extraction scheme that extends QRP to spin-1 systems and incorporates two-body spin-correlation observables into the reservoir readout. Compared with local QRP readouts and conventional state-based features, classifiers trained on CEQRP features exhibit substantially improved cross-family generalization. In particular, logistic regression trained on full-information CEQRP features and AdaBoost trained on ellipsoid-like CEQRP features both achieve high accuracy on previously unseen Werner and isotropic states across multiple measurement settings.

Second, independently of CEQRP, we establish two interconnected steering criteria for mixed two-qutrit states: a necessary and sufficient all-versus-nothing (AVN) criterion for states admitting three linearly independent pure conditional states under Alice's projective measurement, and an asymptotic sufficient criterion based on transverse coherence and population leakage. We show that states in the AVN class can be transformed, through invertible local filtering on Bob's side, into states satisfying the asymptotic criterion. These results enable the construction of structurally distinct, rigorously certified steerable samples for training-set augmentation.

Combining these advances yields our central physical result: the identification of new Alice-to-Bob steerable partially entangled (PE) states. Specifically, classifiers trained on CEQRP features and theory-augmented datasets identify candidate PE states at $p=0.45$, whose steerability is independently confirmed by semidefinite programming (SDP) using seven projective measurement settings on Alice's side. We further use classifier predictions to guide finite-setting SDP searches and investigate the hierarchy of steering detection with different numbers of measurement settings.

Together, these results demonstrate how correlation-enhanced quantum reservoir features and independent theoretical steering criteria can jointly advance generalizable classification and extend the reach of high-dimensional quantum steering detection.

\section{Correlation-Enhanced QRP Learning for Cross-Family Generalization}

The finite-measurement
SDP datasets, introduced in Ref.~\cite{h7qp-63dg}, are denoted by
$
\mathcal{OD}_{m},m=3,4,5,6,7.
$
In these datasets, steering is tested using an SDP with a fixed number of measurement settings and a finite number of optimization iterations. Classifiers trained with either the full-information representation \(F_1\) or
the steering-ellipsoid-like representation \(F_2\) generalize well to isotropic states but achieve only approximately \(50\%\) accuracy on previously unseen Werner states.
These results reveal the limitations of conventional state-based representations in cross-family steering classification. This motivates us to explore whether QRP
dynamics can provide more effective representations
for steering classification.

To this end, we extend QRP from spin-$1/2$ to spin-1
systems, employing a transverse-field Ising reservoir
with single-ion anisotropy \cite{WU20103502},
\begin{equation}
	H^R=
	-J\sum_{\langle i,k\rangle}S_i^zS_k^z
	-h\sum_{i=1}^{N}S_i^x
	+D\sum_{i=1}^{N}(S_i^z)^2,
	\label{eq:spin1_qrp_hamiltonian}
\end{equation}
where $S_i^\alpha$ ($\alpha=x,y,z$) are spin-1 operators, $J$ and $h$ denote the Ising coupling and transverse-field strengths, respectively, and $D$ is the single-ion anisotropy.
Unlike spin-$1/2$, for which $(S_i^z)^2$ is
proportional to the identity, the corresponding
spin-1 operator is nontrivial and contributes
to the reservoir dynamics.

%We initially adopt local readout observables
%$\{S_i^x,S_i^y,S_i^z,(S_i^z)^2\}_{i=1}^{N}$.
%However, the resulting $F_1$-QRP-MLP classifier
%achieves only about $60\%$ accuracy on Werner states,
%while other classifiers remain near $50\%$.
%Thus, extending the reservoir dynamics to spin-1
%systems with local readout alone yields only
%limited improvement in cross-family generalization.
We first consider the local readout set
\[
\left\{S_i^x,S_i^y,S_i^z,(S_i^z)^2\right\}_{i=1}^{N}.
\]
The resulting \(F_1\)-QRP-MLP classifier achieves only approximately \(60\%\) accuracy on Werner states, while the other classifiers remain close to \(50\%\). Thus, extending the reservoir to spin-1 systems with local observables alone provides only limited improvement in cross-family generalization.

To better capture dynamical correlations, we
augment the local readout with bipartite
correlation observables,
\begin{align}
	\mathcal{O}^R
	={}
	\{S_i^x,S_i^y,S_i^z,(S_i^z)^2\}_{i=1}^{N}
	\cup
	\{S_i^zS_k^z,S_i^xS_k^x\}_{1\leq i<k\leq N}.
	\label{eq:correlation_enhanced_readout}
\end{align}
Here, $S_i^zS_k^z$ probes longitudinal
Ising correlations, whereas $S_i^xS_k^x$
provides complementary information about
transverse spin correlations.
Their expectation values at multiple evolution
times form the {\bf CEQRP features.} See Supplemental Material (SM) \ref{app:qrp_benchmarks} for details.

Incorporating CEQRP features substantially improves cross-family generalization. Table~\ref{tab:two_qutrit_results} summarizes the classification accuracies of the \(F_1\)-CEQRP-LR model for different measurement settings. The CV and validation sets comprise random states, whereas WernerTest and isotropicTest assess generalization to Werner and isotropic states, respectively.

\begin{table}[!h]
	\footnotesize
	\caption{the accuracy of \(F_1\)-CEQRP-LR classification}
	\label{tab:two_qutrit_results}
	\begin{ruledtabular}
		\begin{tabular}{cccccc}
			QRCML &  m=3 & m=4 & m=5 & m=6 & m=7 \\
			\hline
			CV & 87.59\% & 76.83\% & 85.58\% & 84.59\% & 84.36\% \\
			Validation & 86.75\% & 76.81\% & 85.78\% & 85.75\% & 85.06\% \\
			WernerTest & 74.95\% & 96.00\% & 96.03\% & 77.45\% & 92.85\%\\
			isotropicTest & 77.83\% & 99.35\% & 97.08\% & 77.08\% & 95.10\%\\
			
	\end{tabular}\end{ruledtabular}
\end{table}

For \(m=4\) and \(m=5\), the classifier achieves \(96.00\%\) and \(96.03\%\) accuracy on unseen Werner states, respectively, together with \(99.35\%\) and \(97.08\%\) accuracy on isotropic states. The \(m=5\) model also achieves higher accuracy on random states, reaching \(85.58\%\) in cross-validation and \(85.78\%\) on the validation set. Thus, both measurement settings yield strong cross-family generalization, with \(m=5\) providing a more balanced performance across the three state families. At \(m=7\), the accuracies on Werner and isotropic states remain high, at \(92.85\%\) and \(95.10\%\), respectively, whereas the Werner-state accuracies at \(m=3\) and \(m=6\) are lower, at \(74.95\%\) and \(77.45\%\).

The improvement also extends to \(F_2\) features. As shown in Table~\ref{tab:F2_two_qutrit_results}, \(F_2\)-CEQRP-AdaBoost achieves Werner-state accuracies of \(73.72\%\)$\sim$\(82.52\%\) and isotropic-state accuracies of \(96.97\%\)$\sim$\(99.65\%\) across the five measurement settings. Its cross-validation and test accuracies on random states range from \(83.75\%\) to \(89.23\%\) and from \(83.93\%\) to \(88.61\%\), respectively. Detailed results for other classifiers and measurement settings are provided in the SM \ref{app:ceqrp_full_readout_results}.

  \begin{table}[!h]
	\footnotesize
	\caption{the accuracy of \(F_2\)-CEQRP-AdaBoost classification on SDP-labeled datasets.}
	\label{tab:F2_two_qutrit_results}
	\begin{ruledtabular}
		\begin{tabular}{cccccc}
			CEQRCML &  m=3 & m=4 & m=5 & m=6 & m=7 \\
			\hline
			crossrate & 89.23\% & 83.75\% & 86.94\% & 86.33\% & 87.31\% \\
			testrate & 88.61\% & 83.93\% & 87.50\% & 88.38\% & 87.65\% \\
			Wernerrate & 73.72\% & 82.52\% & 80.32\% & 82.10\% & 81.17\% \\
			isotropicrate & 99.65\% & 96.97\% & 97.87\% & 97.15\% & 97.52\% \\
		\end{tabular}
	\end{ruledtabular}
\end{table}

For \(F_1\)-CEQRP features, LR outperforms the multilayer perceptron (MLP) at (m=3,4,5), demonstrating that a more complex classifier does not necessarily yield better generalization.

{\it These results suggest that
	CEQRP improves the transferability of learned
	steering classifiers across different
	quantum state families.} Nevertheless, improved feature representations alone do not suffice to identify new steerable states, motivating the incorporation of theory-certified samples into the training set.

\section{Interconnected Steering Criteria and Theory-Certified Data Augmentation}

We establish two interconnected criteria for certifying Alice-to-Bob steering in mixed two-qutrit states. The first extends the AVN approach \cite{ref13} to two-qutrit systems, while the second exploits the asymptotic behavior of conditional states near the boundary
of quantum state space.

We first consider states for which a
projective measurement on Alice's subsystem
prepares three linearly independent pure
conditional states on Bob's subsystem.
Up to a local unitary
transformation, such states can be expressed as
\begin{align}
	\rho_{AB}
	={}&
	\sum_{j=0}^{2}\mu_{j+1}
	|j\rangle\langle j|
	\otimes|\phi_{j+1}\rangle\langle\phi_{j+1}|
	\nonumber\\
	&+
	\bigl(
	|0\rangle\langle1|\otimes\mathcal A
	+
	|0\rangle\langle2|\otimes\mathcal B
	+
	|1\rangle\langle2|\otimes\mathcal C
	+\mathrm{H.c.}
	\bigr),
	\label{eq:theory_pure_conditional}
\end{align}
where $\mu_j>0$, $\sum_j\mu_j=1$, and
$\{|\phi_j\rangle\}_{j=1}^{3}$ are linearly
independent normalized vectors.
The operators $\mathcal A$, $\mathcal B$, and
$\mathcal C$ describe the off-diagonal
coherence blocks.

\noindent\textbf{Theorem 1(AVN criterion).}
\textit{A two-qutrit state of the above form
	is unsteerable under projective measurements from Alice to Bob if and only if
	$\mathcal A=\mathcal B=\mathcal C=0$.}

Thus, any nonzero coherence block certifies Alice-to-Bob steering within this class. Theorem 1 provides a necessary and sufficient criterion, completely characterizing the
steerability of states satisfying its assumptions.

Zhang and Chen~\cite{Zhang} established a geometric steering criterion based on first-order transverse coherence and second-order inward deviation, and extended their analysis to general finite-dimensional systems through a support-kernel criterion. Motivated by this geometric perspective, we derive an explicit sufficient criterion for two-qutrit states in terms of asymptotic bounds on transverse coherence and population leakage.  This formulation allows us to construct theory-certified steerable samples for augmenting the CEQRP training dataset.

Let \(\rho_{AB}\) be a two-qutrit state and consider a family of positive measurement operators \(\{M_t\}_{t\in(0,t_0)}\) on Alice's subsystem, satisfying \(0\preceq M_t\preceq I_A\). Put
\begin{equation}\label{MMMM}
\mathsf M_{t}=\{M_{1|t},M_{0|t}\}, M_{1|t}=M_t, M_{0|t}=I_A-M_t.\end{equation}
The corresponding unnormalized conditional states on Bob's subsystem are given by
\begin{eqnarray}
\sigma_t:=\sigma_{1|t}&=&\mathrm{Tr}_{A}\left[(M_{1|t}\otimes I_B)\rho_{AB}\right],\label{SSSS-1}\\ \sigma_{0|t}&=&\mathrm{Tr}_{A}\left[(M_{0|t}\otimes I_B)\rho_{AB}\right].\label{SSSS-0}
\end{eqnarray}
In a fixed orthonormal basis \(\{|0\rangle,|1\rangle,|2\rangle\}\), we write
\begin{equation}
	\sigma_t
	=
	\begin{pmatrix}
		a_t & b_t & c_t\\
		\overline{b_t} & d_t & e_t\\
		\overline{c_t} & \overline{e_t} & f_t
	\end{pmatrix}.
	\label{eq:sigma_t_matrix}
\end{equation}
We define the transverse coherence \(T_t\) and the longitudinal population leakage \(D_t\) as
\begin{equation}
	T_t
	=
	\sqrt{3\left(|b_t|^2+|c_t|^2\right)},
	\qquad
	D_t=d_t+f_t.
\end{equation}
These quantities characterize the coherence between the reference state \(|0\rangle\) and its orthogonal subspace, and the population outside the reference state, respectively.

\noindent\textbf{Theorem 2 (Geometry criterion).}
\label{thm:asymptotic_steering} {\it Suppose that there exist constants \(L>0\), \(C>0\), and \(t_1\in(0,t_0)\) such that, for all \(0<t<t_1\),
	\begin{equation}
		T_t\geq Lt,
		\qquad
		D_t\leq Ct^2.
	\end{equation}
	Then the family of conditional states \(\{\sigma_t:0<t<t_1\}\) does not admit a common LHS model, i.e., there does not exist a probability measure space \((\Lambda,\Sigma,\mu)\),  a measurable family of density operators \(\{\tau_\lambda\}_{\lambda\in\Lambda}\),  and measurable response functions \(P_t:\Lambda\to[0,1]\) such that
	\begin{equation}\label{ZTHM2}
		\sigma_t
		=
		\int_{\Lambda}
		P_t(\lambda)\tau_\lambda\,d\mu(\lambda),
		\qquad
		\forall 0<t<t_1.
		%\label{eq:lhs-model}
	\end{equation} Consequently, the corresponding two-qutrit state \(\rho_{AB}\) is steerable from Alice to Bob under finitely many of two-outcome POVM measurements  $\mathsf M_{t_1},\mathsf M_{t_2},\ldots,\mathsf M_{t_m}$ defined by \eqref{MMMM}.}

Theorem 2 establishes a sufficient steering
criterion through the contrasting asymptotic
bounds on transverse coherence and
population leakage.

The two criteria are connected through
invertible local filtering on Bob's
subsystem. Since the conditional states
in Eq.~(\ref{eq:theory_pure_conditional}) are linearly
independent, the matrix
$
\Phi=
\bigl(
|\phi_1\rangle,
|\phi_2\rangle,
|\phi_3\rangle
\bigr)
$
is invertible.
An appropriate local filter
$K=\eta\Phi^{-1}$, with
$K^\dagger K\preceq I_B$,
transforms the state into
\begin{align}
	\widetilde{\rho}_{AB}
	={}&
	\sum_{j=0}^{2}\tilde{\mu}_{j+1}
	|j\rangle\langle j|
	\otimes|j\rangle\langle j|
	\nonumber\\
	&+
	\bigl(
	|0\rangle\langle1|\otimes\tilde{\mathcal A}
	+
	|0\rangle\langle2|\otimes\tilde{\mathcal B}
	+
	|1\rangle\langle2|\otimes\tilde{\mathcal C}
	+\mathrm{H.c.}
	\bigr),
	\label{eq:theory_pure_conditional-1}
\end{align}
whose conditional pure states are orthonormal.
Because the filter is invertible
and acts on the trusted subsystem,
Alice-to-Bob steerability is preserved.

For states of the form (\ref{eq:theory_pure_conditional-1}), whose ranks are less than $4$, the positive partial transpose (PPT) criterion provides a convenient test for entanglement \cite{PRA.62.032310}. If they are entangled, then they satisfy the assumption in Theorem 2.  Consequently, every steerable AVN state covered by Theorem 1 can be transformed, through an invertible local filter on Bob's subsystem, into a state satisfying Theorem 2.
Theorem 2 also certifies mixed states with structures not captured by the AVN form.
These results connect the exact coherence-based characterization with a broader asymptotic approach
to steering certification. Detailed proofs and examples are provided in the SM \ref{app:thm_proof}-\ref{app:sec:connection_theorems}.

Beyond their theoretical significance, the two criteria provide rigorously certified steerable samples for
supervised learning. Specifically, we generate orthogonal
AVN states satisfying Theorem 1 and additional mixed states certified by Theorem 2. These samples augment the original
dataset $\mathcal{D}_{\mathrm{acc}}$ to construct $\mathcal{D}_{\mathrm{cert}}$. The detailed generation procedure is provided in the SM \ref{app:theorem_state_families}.

This augmentation introduces physically certified states with different structural properties. Combined with CEQRP, it enables the identification of
new steerable PE states, as demonstrated below.

\section{Discovering New Steerable Partially Entangled States}

We now investigate whether combining CEQRP with theory-certified data augmentation can identify new steerable PE states in two-qutrit systems.

The PE-state family is defined as
\begin{equation}
	\rho_{\mathrm{PE}}(p,\theta,\phi)
	=
	p|\psi_{\theta,\phi}\rangle
	\langle\psi_{\theta,\phi}|
	+
	(1-p)\rho_A\otimes\frac{I_3}{3},
	\label{eq:PE_states}
\end{equation}
where
$|\psi_{\theta,\phi}\rangle
=
\cos\theta\sin\phi\,|00\rangle
+\sin\theta\sin\phi\,|11\rangle
+\cos\phi\,|22\rangle$,
and
$\rho_A=\operatorname{Tr}_B
(|\psi_{\theta,\phi}\rangle
\langle\psi_{\theta,\phi}|)$.

The PE-state family exhibits directional steering
asymmetry. For steering from Bob to Alice,
an analytical nonsteerability bound
$p\leq 5/12$ was established for projective
measurements, while four mutually unbiased
bases (MUBs) can detect steering for
$p>0.4818$ \cite{ref33}.

For the opposite direction, Ref.\cite{ref33}
employed a shrinking-factor-assisted SDP
method to obtain parameter-dependent sufficient
bounds for Alice-to-Bob nonsteerability
under projective measurements.
The resulting bounds vary approximately
from $0.1$ to $1$ with $(\theta,\phi)$.
Together, these results establish one-way
steering in certain parameter regions,
but do not fully characterize the
Alice-to-Bob steering boundary.

Here, we investigate whether theory-certified data augmentation can help identify new Alice-to-Bob steerable PE states. We first examine PE states using the accurately labeled dataset \(\mathcal D_{\mathrm{acc}}\) of Ref.~\cite{h7qp-63dg}, which contains steerable Werner and isotropic states, entangled pure states, SDP-certified steerable random states, and known unsteerable states. In the parameter range examined, steering-weight (SW) analysis does not certify Alice-to-Bob steering for PE states with \(p<0.4818\). Similarly, conventional ML and CEQRP classifiers trained on \(\mathcal D_{\mathrm{acc}}\), without theory-certified data augmentation, fail to identify such states.

\begin{figure}[!h]
	\centering
	\includegraphics[width=0.9\columnwidth]
	{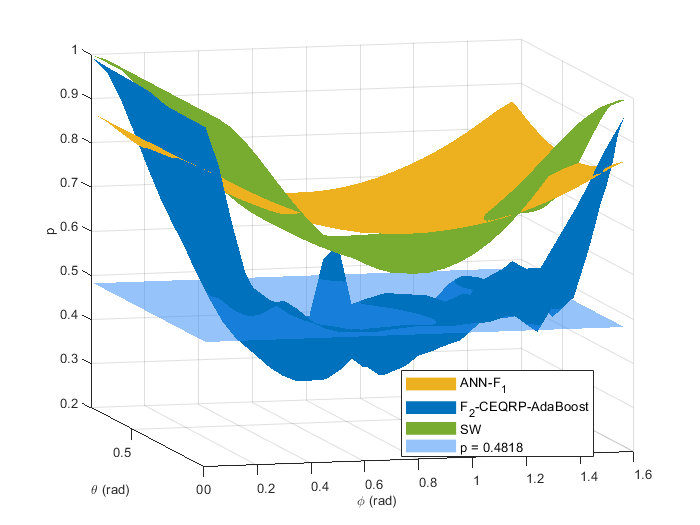}
	\caption{
		Predicted Alice-to-Bob steering boundaries
		for two-qutrit PE states using CEQRP-based
		classifiers.
		The predictions identify candidate
		steerable regions at $p<0.4818$.
		This reference value originates from
		four-MUB Bob-to-Alice steering detection,
		rather than an Alice-to-Bob threshold.
	}
	\label{fig:pe-bounds}
\end{figure}

\textbf{Discoveries:} \textit{Augmenting \(\mathcal D_{\mathrm{acc}}\) with states certified by Theorems 1 and 2 yields the dataset \(\mathcal D_{\mathrm{cert}}\). An AdaBoost classifier trained on \(F_2\)-CEQRP features then identifies candidate Alice-to-Bob steerable regions for \(p<0.4818\), as shown in Fig.~\ref{fig:pe-bounds}.}

To verify the CEQRP predictions in Discoveries independently, we fix $\theta=\pi/4$ and examine the Alice-to-Bob steering of the predicted PE states using SDP with seven projective
measurement settings on Alice. The SDP tests whether the corresponding conditional states admit a LHS model.

\begin{figure}[!htbp]
	\centering
	\includegraphics[width=\columnwidth]{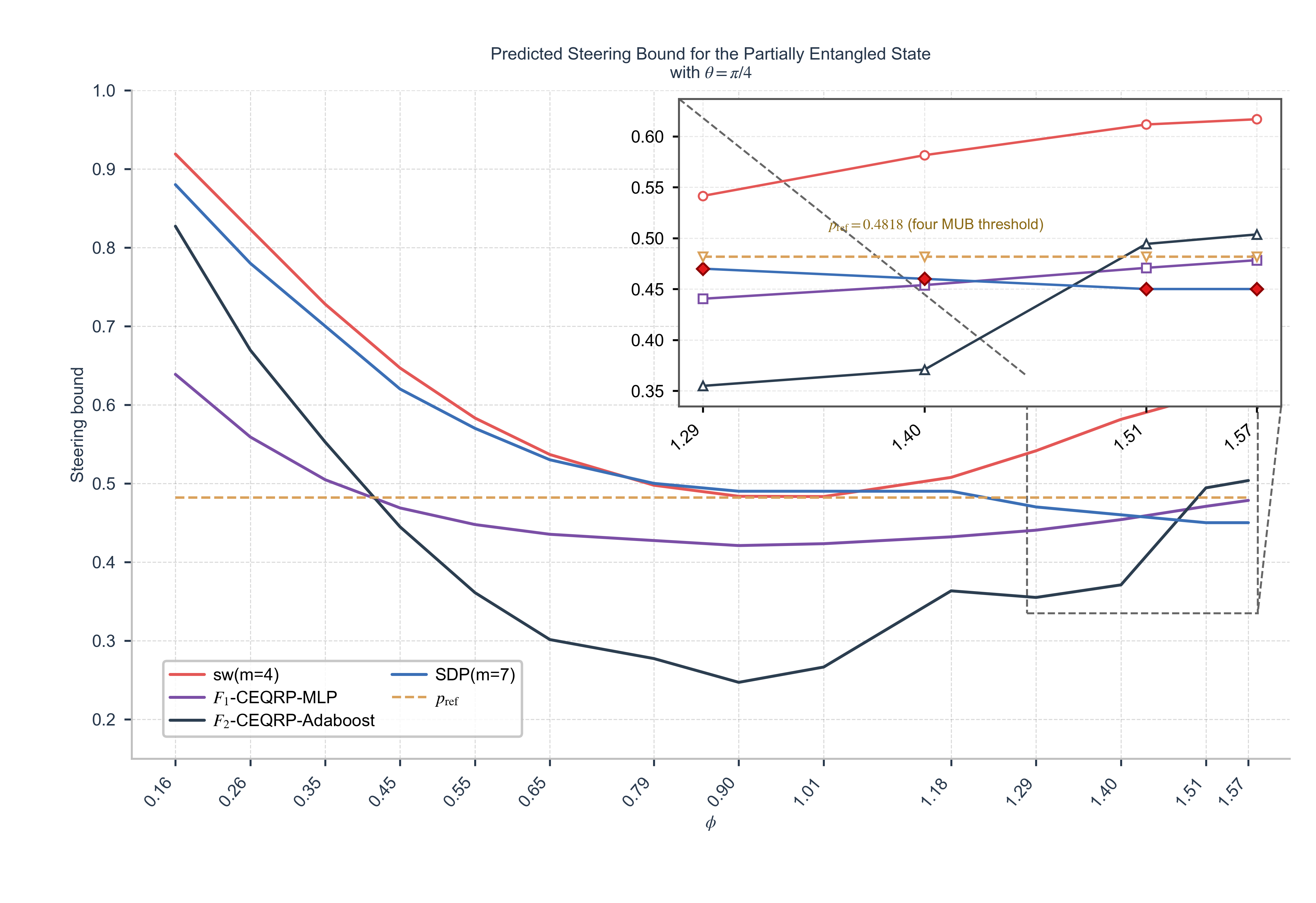}
	\caption{
		Alice-to-Bob steering detection for two-qutrit PE states
		with $\theta=\pi/4$. The $F_2$-CEQRP-AdaBoost predictions
		are compared with previous ML and steering-weight results
		and independent SDP calculations using seven projective
		measurements. SDP certifies steering below $p=0.4818$; the
		inset highlights the newly identified steerable region.
	}
	\label{fig:theta-pi4-qrc-sw}
\end{figure}

Remarkably, the SDP calculations detect
Alice-to-Bob steering for
\begin{align}
	(\phi,p)
	={}&(1.2871,0.47),\ (1.3428,0.46),
	\nonumber\\
	&(1.4543,0.45),\ (1.5101,0.45).
	\label{eq:new_PE_steerable_states}
\end{align}
These results establish the existence of new Alice-to-Bob steerable PE states with \(p<0.4818\). Figure~\ref{fig:theta-pi4-qrc-sw} compares the steering boundaries predicted by CEQRP-based classifiers with finite-setting SDP results and previous benchmarks. The inset highlights the newly identified steerable region, with SDP-certified points explicitly marked to distinguish rigorous certification from classifier predictions.
%\begin{figure*}[!t]
%	\centering
%	\includegraphics[width=0.8\textwidth]{./PESB_PI4.png}
%	\caption{
%		Alice-to-Bob steering detection for
%		two-qutrit PE states with $\theta=\pi/4$.
%		The $F_2$-CEQRP-AdaBoost predictions are
%		compared with previous ML and
%		steering-weight results and independent
%		SDP calculations using seven projective
%		measurement settings.
%		The SDP results certify steering at
%		selected parameter values below
%		$p=0.4818$. The inset highlights the newly identified
%		steerable region.
%	}
%	\label{fig:theta-pi4-qrc-sw}
%\end{figure*}

%These findings extend the known Alice-to-Bob steerable region of two-qutrit PE states and highlight the role of theory-certified QRP learning in discovering previously uncertified quantum steering.
The SDP certificates extend the set of known Alice-to-Bob steerable PE states to selected points with \(p<0.4818\). These results demonstrate that theory-certified data augmentation can guide CEQRP-based searches for steerable states, with the SDP providing independent certification.

\section{Exploring the Measurement-Setting Hierarchy with CEQRP Learning}
\label{sec:measurement_hierarchy}

Beyond establishing whether a state is steerable, it is important to determine the number of measurement settings required to reveal its steering. A state undetected with a limited number of settings may become detectable when
additional settings are allowed, giving rise to a measurement-setting hierarchy of quantum steering.
Such a hierarchy was investigated for two-qubit systems by Wang \textit{et al.}~\cite{ref29}. Here, we explore its extension to two-qutrit PE states using CEQRP-based
learning.

\begin{figure}[!h]
	\centering
	\includegraphics[width=\columnwidth]
	{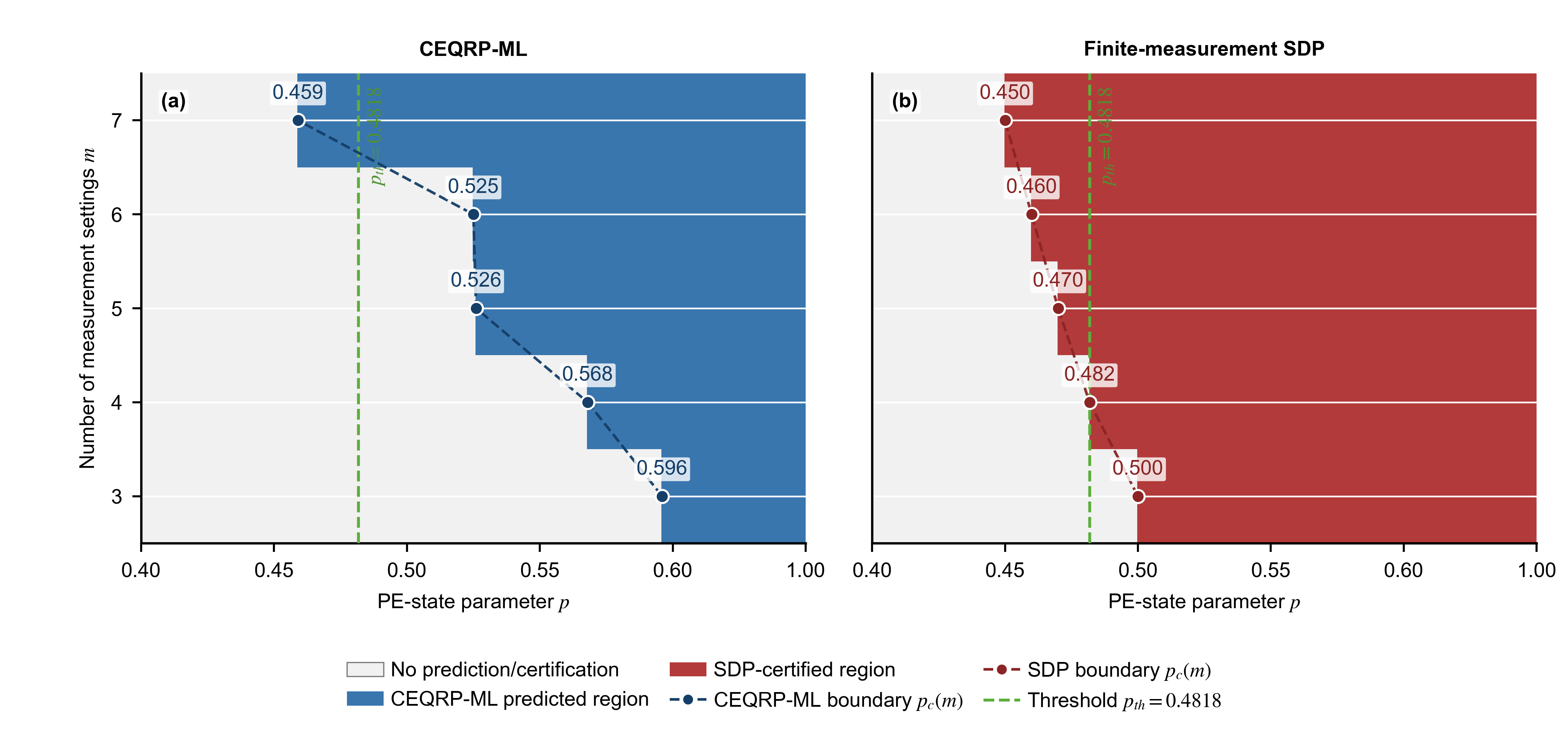}
	\caption{
		Measurement-setting hierarchy of
		Alice-to-Bob steering detection for
		two-qutrit PE states with
		$\theta=\pi/4$ and $\phi=1.5101$.
		The horizontal axis denotes the state
		parameter $p$, and the vertical axis
		represents the number of projective
		measurement settings $m=3,\ldots,7$.
		(a) Steering regions predicted by
		CEQRP-based classifiers.
		(b) Regions detected by independent
		finite-setting SDP searches with
		600 optimization iterations.
	}
	\label{fig:pe_measurement_number}
\end{figure}

We consider the PE-state family with
$\theta=\pi/4$ and $\phi=1.5101$,
for which Alice-to-Bob steering was
certified at $p=0.45$ using seven
projective measurement settings.
To investigate the dependence of
steering detection on the number of
settings $m=3,\ldots,7$, we train
classifiers on the corresponding
datasets $\mathcal{OD}_m$, using
$F_1$- and
$F_2$-CEQRP features.
For each $m$, the classifiers yield
an estimated detection boundary
$\widehat p_{\mathrm c}^{\mathrm{QRP}}(m)$.
The predicted steering regions are
shown in Fig.~\ref{fig:pe_measurement_number}(a).

We test these predictions independently by performing finite-setting SDP searches over projective measurements. Each search uses \(600\) optimization iterations, with the CEQRP-based predictions determining the parameter intervals to be examined. \cite{ref45}
%The numerical search employs 600
%optimization iterations, with the QRP
%predictions used to guide the choice
%of parameter regions.
For example, the CEQRP-based model predicts
$\widehat p_{\mathrm c}^{\mathrm{QRP}}(3)
\simeq 0.596$.
We therefore initialize the three-setting SDP search near $p=0.6$ and investigate smaller values of $p$ to locate the
numerical detection boundary.

The search also exploits the measurement
hierarchy: any steering demonstration
using $m$ settings remains valid when
additional settings are allowed.
Consequently, the optimal detection
thresholds obey
\begin{equation}
	p_{\mathrm c}^{\mathrm{opt}}(3)
	\geq p_{\mathrm c}^{\mathrm{opt}}(4)
	\geq \cdots
	\geq p_{\mathrm c}^{\mathrm{opt}}(7).
	\label{eq:measurement_hierarchy}
\end{equation}
This ordering, together with the steering boundaries predicted by the CEQRP-based classifiers, narrows the parameter intervals used in the SDP searches. The resulting finite-setting SDP regions are shown in Fig.~\ref{fig:pe_measurement_number}(b).
%This relation, together with the
%QRP-predicted boundaries, helps
%identify parameter regions for
%targeted SDP searches.
%The corresponding SDP-detected regions
%are presented in
%Fig.~\ref{fig:pe_measurement_number}(b).

The comparison indicates that CEQRP-based learning captures how the detected steering region depends on the measurement budget. Moreover, the CEQRP-based predictions supply useful initial intervals for finite-setting SDP searches, thereby providing a practical approach to exploring measurement-dependent steering boundaries in higher-dimensional quantum systems.

\section{Conclusion and Discussion}

Our results demonstrate the potential of quantum reservoir
learning to move beyond the classification of known
steering correlations toward the discovery of previously
uncertified steerable states. CEQRP provides features that
substantially improve cross-family generalization,
while independent steering criteria supply rigorously
certified samples for training-set augmentation.
Combining these advances, CEQRP-based classifiers identify
new Alice-to-Bob steerable partially
PE states, including examples at $p=0.45$
independently verified by SDP.

These findings raise several interconnected questions.
Which two-body correlations in the CEQRP readout
are responsible for improved cross-family
generalization, and how do they contribute to
identifying steerable PE states?
Can CEQRP-guided exploration, combined with
further developments of the theoretical steering
criteria, lead to a more complete characterization
of the steering regions of PE states?
More importantly, can any of the newly identified
PE states be shown to be steerable from Alice to Bob
using projective measurements, while remaining
unsteerable from Bob to Alice under all projective
measurements?
Such a result would establish one-way steering
in the direction opposite to the previously reported
Bob-to-Alice-steerable and Alice-to-Bob-unsteerable
regime for this family.
Addressing these questions may deepen our understanding
of the interplay between quantum reservoir features
and theoretical steering certification, while
revealing new forms of directional steering in
high-dimensional quantum systems.

\section*{Acknowledgments}
	Z. Y. Li was supported by the National Natural Science Foundation of China (Grant No. 12571560). H. X. Meng was supported by the Beijing Natural Science Foundation (Grant No. 1262039) and the National Natural Science Foundation of China (Grant No. 12461087).

%\begin{acknowledgments}
%
%
%\end{acknowledgments}

\nocite{*}

\bibliography{ref}

\clearpage
\appendix
\setcounter{page}{1}
\renewcommand{\thesubsection}{S\arabic{subsection}}
\renewcommand{\thesubsection}{S\arabic{subsection}}
\setcounter{table}{0}
\renewcommand{\thetable}{\arabic{table}}
\setcounter{figure}{0}
\renewcommand{\thefigure}{\arabic{figure}}
\setcounter{equation}{0}
\renewcommand{\theequation}{\arabic{equation}}
\onecolumngrid

\section*{Supplementary Material}
\setcounter{subsection}{0}

\subsection{CEQRP Features}
\label{app:qrp_benchmarks}

This subsection introduces: the definition of the correlation-enhanced quantum reservoir processing (CEQRP) feature vector that is used throughout the main text.

\paragraph{Input-state representations.}
We consider the two input representations \(F_1\) and \(F_2\) introduced in Ref.~\cite{h7qp-63dg}. The full-information representation \(F_1\) uses the original state \(\rho_{AB}\), whereas the compact representation \(F_2\) uses a transformed state
\(\rho'_{AB}\) whose construction is detailed in
Ref.~\cite{h7qp-63dg}. The transformation defining \(F_2\) preserves Alice-to-Bob steerability. We denote the physical input state associated with representation \(j\) by
\begin{equation}
	\widetilde{\rho}_{AB}^{(1)}=\rho_{AB},
	\qquad
	\widetilde{\rho}_{AB}^{(2)}=\rho'_{AB},
	\qquad j\in\{1,2\}.
	\label{eq:app_ceqrp_input_representations}
\end{equation}
Thus, \(F_1\) and \(F_2\) specify which state is supplied to the
reservoir; the CEQRP feature vector is then generated from the
reservoir response to that input state.

\paragraph{Reservoir initialization and evolution.}
The two qutrits \(A\) and \(B\) are encoded into the first two sites
of an \(N\)-site spin-1 reservoir. The remaining \(N-2\) sites are
initialized in the same reference state \(\ket{0}\), where
\(\ket{0}\) denotes the single-site reservoir state used in the
simulations. The initial reservoir state for representation \(j\)
is therefore
\begin{equation}
	\rho^R_{(j)}(0)
	=
	\widetilde{\rho}_{AB}^{(j)}
	\otimes
	(\ket{0}\bra{0})^{\otimes(N-2)},
	\qquad j\in\{1,2\}.
	\label{eq:app_ceqrp_initial_state}
\end{equation}

The reservoir evolves under the spin-1 transverse-field Ising
Hamiltonian with single-ion anisotropy,
\begin{equation}
	H^R
	=
	-J\sum_{\langle i,k\rangle}S_i^zS_k^z
	-h\sum_{i=1}^{N}S_i^x
	+D\sum_{i=1}^{N}(S_i^z)^2,
	\label{eq:app_ceqrp_hamiltonian}
\end{equation}
where \(S_i^\alpha\) (\(\alpha=x,y,z\)) are spin-1 operators acting
on site \(i\). The parameters \(J\), \(h\), and \(D\) denote the
Ising coupling, transverse-field strength, and single-ion
anisotropy, respectively. These parameters, together with the
coupling configuration, are fixed for all input states and
throughout feature extraction; their values are drawn from
\begin{equation}
	J\in\{-1,-0.5,0.5,1\},
	\qquad
	h\in[0,1.5],
	\qquad
	D\in[-1.5,1.5],
	\label{eq:app_ceqrp_parameter_ranges}
\end{equation}
and are kept identical for every sample used in a given numerical
experiment. We work in units with \(\hbar=1\), so that the
reservoir state at time \(t\) is
\begin{equation}
	\rho^R_{(j)}(t)
	=
	U^R(t)\,\rho^R_{(j)}(0)\,\bigl(U^R(t)\bigr)^\dagger,
	\qquad
	U^R(t)=e^{-iH^Rt}.
	\label{eq:app_ceqrp_evolution}
\end{equation}

\paragraph{Readout observables.}
The CEQRP readout includes both local spin observables and two-site
correlation observables. The local readout set is
\begin{equation}
	\mathcal{O}_{\mathrm{loc}}^R
	=
	\left\{
	S_i^x,S_i^y,S_i^z,(S_i^z)^2
	\right\}_{i=1}^{N}.
	\label{eq:app_ceqrp_local_readout}
\end{equation}
To supplement these single-site measurements with information about
inter-site correlations, we add the observables
\(S_i^zS_k^z\) and \(S_i^xS_k^x\) for every pair \(i<k\). The
complete CEQRP readout set is
\begin{equation}
	\mathcal{O}^R
	=
	\mathcal{O}_{\mathrm{loc}}^R
	\cup
	\left\{
	S_i^zS_k^z,\;S_i^xS_k^x
	\right\}_{1\leq i<k\leq N}.
	\label{eq:app_ceqrp_readout}
\end{equation}
Here, \(S_i^zS_k^z\) probes longitudinal Ising correlations, while
\(S_i^xS_k^x\) probes transverse spin correlations. Since the
reservoir is spin 1, the local observable \((S_i^z)^2\) is also
nontrivial and contributes to the readout.

The number of observables in \(\mathcal{O}^R\) is
\begin{equation}
	M
	=
	4N+2\binom{N}{2}
	=
	N(N+3).
	\label{eq:app_ceqrp_number_of_observables}
\end{equation}
The first term counts the four local observables per site, and the
second counts the two correlation observables for each distinct
site pair.

\paragraph{Time sampling and feature-vector construction.}
For each observable \(O_\mu^R\in\mathcal{O}^R\), its response for
input representation \(j\) at time \(t_\ell\) is defined by
\begin{equation}
	x_{\mu\ell}^{(j)}
	=
	\operatorname{Tr}
	\!\left[
	O_\mu^R\,\rho^R_{(j)}(t_\ell)
	\right].
	\label{eq:app_ceqrp_expectation_values}
\end{equation}
The expectation values are evaluated at the sampling times
\begin{equation}
	\mathcal{T}
	=
	\{0.50,\,0.75,\,1.00,\,1.50,\,2.00\},
	\qquad
	L=|\mathcal{T}|=5.
	\label{eq:app_ceqrp_sampling_times}
\end{equation}
For a given input representation \(j\), we arrange these responses
into an \(M\times L\) feature matrix
\begin{equation}
	\mathbf{X}^{(j)}
	=
	\left[
	x_{\mu\ell}^{(j)}
	\right]_{\substack{1\leq\mu\leq M\\1\leq\ell\leq L}}
	\in\mathbb{R}^{M\times L}.
	\label{eq:app_ceqrp_feature_matrix}
\end{equation}
The CEQRP feature vector is obtained by stacking the columns of this
matrix:
\begin{equation}
	\boldsymbol{x}_{\mathrm{CEQRP}}^{(j)}
	=
	\operatorname{vec}\!\left(\mathbf{X}^{(j)}\right)
	=
	\left(
	x_{11}^{(j)},\ldots,x_{M1}^{(j)},
	x_{12}^{(j)},\ldots,x_{M2}^{(j)},
	\ldots,
	x_{1L}^{(j)},\ldots,x_{ML}^{(j)}
	\right)^{\mathsf T}.
	\label{eq:app_ceqrp_feature_vector}
\end{equation}
Thus, each component of the feature vector is the expectation value
of one readout observable at one sampling time. Combining
Eqs.~\eqref{eq:app_ceqrp_number_of_observables}
and~\eqref{eq:app_ceqrp_sampling_times}, the feature-vector
dimension grows with the reservoir size as
\begin{equation}
	\dim\!\left(\boldsymbol{x}_{\mathrm{CEQRP}}^{(j)}\right)
	=
	ML
	=
	5N(N+3),
	\label{eq:app_ceqrp_feature_dimension}
\end{equation}
which is the dimension of the feature space used by every classical readout model in~\ref{app:ceqrp_full_readout_results}.

The resulting vector is the CEQRP feature representation of the
input state. It is constructed entirely from reservoir expectation
values and is subsequently supplied to a classical readout model.

\subsection{Classification using the correlation-enhanced readout}
\label{app:ceqrp_full_readout_results}

This section reports the training and testing results of the \(F_i\)-QRP-ML and \(F_i\)-CEQRP-ML classifiers on the SDP-labeled datasets \(\mathcal{OD}_m\). The \(F_i\)-QRP-ML classifiers use only the local readout set \(\mathcal{O}_{\mathrm{loc}}^R\), whereas the \(F_i\)-CEQRP-ML classifiers use the full readout set \(\mathcal{O}^R\) in Eq.~\eqref{eq:app_ceqrp_readout}, which combines local observables with two-site correlations. The CEQRP readout exhibits markedly stronger cross-family generalization than the local-readout-only approach.

We use the same reservoir dynamics, sampling times, input representations, and data protocol for both readout choices. The resulting feature vectors are supplied to the following classical readout models: logistic regression (LR), support vector machines
(SVMs), artificial neural networks (ANNs), gradient boosting (GB), AdaBoost, an ensemble model (EM), and the broad residual learning system (BRLS)~\cite{SU2026112134}. Within each readout choice, all classifiers use the same features; only the classical readout model and its parameters vary.

The classifiers are trained using the SDP-labeled two-qutrit datasets \(\mathcal{OD}_m\) of Ref.~\cite{h7qp-63dg}, where
\(m=3,\ldots,7\) denotes the number of measurement settings. For each \(m\), the \(F_i\)-QRP-ML and \(F_i\)-CEQRP-ML classifiers are trained separately. Stratified cross-validation is used for model selection, and the selected models are evaluated on a
held-out validation set.

Werner states and isotropic states are used as independent structured test families and are excluded from the training set. These tests assess
whether classifiers trained on the random-state datasets generalize
to state families with different structures. Steerable and unsteerable states are assigned the labels \(-1\) and \(+1\),
respectively. All training, validation, and test sets are class-balanced, with equal numbers of steerable and unsteerable samples in each set.

Tables~\ref{tab23}--\ref{tab27} report the classification accuracies of \(F_1\)-CEQRP-ML for \(m=3,\ldots,7\), while Tables~\ref{tab28}--\ref{tab32} report the corresponding results for \(F_2\)-CEQRP-ML. Tables~\ref{tab40}--\ref{tab44} report the classification accuracies of \(F_1\)-CEQRP-ML for \(m=3,\ldots,7\), while Tables~\ref{tab45}--\ref{tab49} report the corresponding results for \(F_2\)-CEQRP-ML.

% ================= 第 1 组：tab23 + tab24 =================
\begin{table}[!htbp]
	\centering
	\begin{minipage}[t]{0.49\textwidth}
		\centering
		\scriptsize
		\setlength{\tabcolsep}{1pt}
		\caption{Classification results of \(F_1\)-CEQRP-ML (\(\mathcal{OD}_3\)).}
		\label{tab23}
		\begin{tabular}{lccccccc}
			\toprule
			Models & LR & SVM & ANN & GB & AdaBoost & EM & BRLS \\
			\midrule
			CV           & 0.8759 & 0.8670 & 0.8650 & 0.8630 & 0.8433 & 0.8761 & 0.7977 \\
			Validation   & 0.8675 & 0.8575 & 0.8619 & 0.8519 & 0.8206 & 0.8625 & 0.8087 \\
			WernerTest   & 0.7495 & 0.5000 & 0.7033 & 0.5000 & 0.5000 & 0.5000 & 0.7872 \\
			IsotropicTest& 0.7783 & 0.5000 & 0.7413 & 0.8853 & 0.8715 & 0.6743 & 0.5000 \\
			\bottomrule
		\end{tabular}
	\end{minipage}\hfill
	\begin{minipage}[t]{0.49\textwidth}
		\centering
		\scriptsize
		\setlength{\tabcolsep}{1pt}
		\caption{Classification results of \(F_1\)-CEQRP-ML (\(\mathcal{OD}_4\)).}
		\label{tab24}
		\begin{tabular}{lccccccc}
			\toprule
			Models & LR & SVM & ANN & GB & AdaBoost & EM & BRLS \\
			\midrule
			CV           & 0.7683 & 0.7494 & 0.7380 & 0.7577 & 0.7334 & 0.7650 & 0.7089 \\
			Validation   & 0.7681 & 0.7475 & 0.7544 & 0.7675 & 0.7469 & 0.7694 & 0.7019 \\
			WernerTest   & 0.9600 & 0.5000 & 0.9685 & 0.5000 & 0.5000 & 0.5000 & 0.5243 \\
			IsotropicTest& 0.9935 & 0.5000 & 0.9493 & 0.5000 & 0.8643 & 0.5000 & 0.9610 \\
			\bottomrule
		\end{tabular}
	\end{minipage}
\end{table}

% ================= 第 2 组：tab25 + tab26 =================
\begin{table}[!htbp]
	\centering
	\begin{minipage}[t]{0.49\textwidth}
		\centering
		\scriptsize
		\setlength{\tabcolsep}{1pt}
		\caption{Classification results of \(F_1\)-CEQRP-ML (\(\mathcal{OD}_5\)).}
		\label{tab25}
		\begin{tabular}{lccccccc}
			\toprule
			Models & LR & SVM & ANN & GB & AdaBoost & EM & BRLS \\
			\midrule
			CV           & 0.8558 & 0.8434 & 0.8380 & 0.8308 & 0.8045 & 0.8455 & 0.7794 \\
			Validation   & 0.8578 & 0.8431 & 0.8475 & 0.8325 & 0.8075 & 0.8494 & 0.7950 \\
			WernerTest   & 0.9603 & 0.5000 & 0.9480 & 0.5000 & 0.5000 & 0.9415 & 0.5000 \\
			IsotropicTest& 0.9708 & 0.5000 & 0.9125 & 0.8895 & 0.8718 & 0.8588 & 0.5000 \\
			\bottomrule
		\end{tabular}
	\end{minipage}\hfill
	\begin{minipage}[t]{0.49\textwidth}
		\centering
		\scriptsize
		\setlength{\tabcolsep}{1pt}
		\caption{Classification results of \(F_1\)-CEQRP-ML (\(\mathcal{OD}_6\)).}
		\label{tab26}
		\begin{tabular}{lccccccc}
			\toprule
			Models & LR & SVM & ANN & GB & AdaBoost & EM & BRLS \\
			\midrule
			CV           & 0.8459 & 0.8339 & 0.8316 & 0.8280 & 0.8025 & 0.8432 & 0.7794 \\
			Validation   & 0.8575 & 0.8388 & 0.8313 & 0.8319 & 0.8156 & 0.8500 & 0.7744 \\
			WernerTest   & 0.7745 & 0.5000 & 0.9105 & 0.5000 & 0.5000 & 0.5000 & 0.5000 \\
			IsotropicTest& 0.7708 & 0.5000 & 0.8850 & 0.5000 & 0.8713 & 0.5000 & 0.9628 \\
			\bottomrule
		\end{tabular}
	\end{minipage}
\end{table}

% ================= 第 3 组：tab27 + tab28 =================
\begin{table}[!htbp]
	\centering
	\begin{minipage}[t]{0.49\textwidth}
		\centering
		\scriptsize
		\setlength{\tabcolsep}{1pt}
		\caption{Classification results of \(F_1\)-CEQRP-ML (\(\mathcal{OD}_7\)).}
		\label{tab27}
		\begin{tabular}{lccccccc}
			\toprule
			Models & LR & SVM & ANN & GB & AdaBoost & EM & BRLS \\
			\midrule
			CV           & 0.8436 & 0.8825 & 0.9028 & 0.8504 & 0.7972 & 0.8906 & 0.7969 \\
			Validation   & 0.8506 & 0.8956 & 0.9150 & 0.8569 & 0.7794 & 0.9081 & 0.8219 \\
			WernerTest   & 0.9285 & 0.5000 & 0.7745 & 0.5000 & 0.5000 & 0.5000 & 0.5000 \\
			IsotropicTest& 0.9510 & 0.5000 & 0.8770 & 0.5000 & 0.8723 & 0.5000 & 0.9680 \\
			\bottomrule
		\end{tabular}
	\end{minipage}\hfill
	\begin{minipage}[t]{0.49\textwidth}
		\centering
		\scriptsize
		\setlength{\tabcolsep}{1pt}
		\caption{Classification results of \(F_2\)-CEQRP-ML (\(\mathcal{OD}_3\)).}
		\label{tab28}
		\begin{tabular}{lccccccc}
			\toprule
			Models & LR & SVM & ANN & GB & AdaBoost & EM & BRLS \\
			\midrule
			CV           & 0.9218 & 0.9183 & 0.9146 & 0.9129 & 0.8923 & 0.9201 & 0.9231 \\
			Validation   & 0.9196 & 0.9164 & 0.9111 & 0.9089 & 0.8861 & 0.9161 & 0.9243 \\
			WernerTest   & 0.5000 & 0.5000 & 0.5000 & 0.5000 & 0.7372 & 0.5000 & 0.5000 \\
			IsotropicTest& 0.5000 & 0.5000 & 0.8035 & 0.5000 & 0.9965 & 0.7949 & 0.8955 \\
			\bottomrule
		\end{tabular}
	\end{minipage}
\end{table}

% ================= 第 4 组：tab29 + tab30 =================
\begin{table}[!htbp]
	\centering
	\begin{minipage}[t]{0.49\textwidth}
		\centering
		\scriptsize
		\setlength{\tabcolsep}{1pt}
		\caption{Classification results of \(F_2\)-CEQRP-ML (\(\mathcal{OD}_4\)).}
		\label{tab29}
		\begin{tabular}{lccccccc}
			\toprule
			Models & LR & SVM & ANN & GB & AdaBoost & EM & BRLS \\
			\midrule
			CV           & 0.8676 & 0.8671 & 0.8661 & 0.8633 & 0.8375 & 0.8697 & 0.8696 \\
			Validation   & 0.8708 & 0.8635 & 0.8658 & 0.8673 & 0.8393 & 0.8675 & 0.8705 \\
			WernerTest   & 0.5000 & 0.5000 & 0.5000 & 0.5000 & 0.8252 & 0.7309 & 0.9800 \\
			IsotropicTest& 0.5000 & 0.5000 & 0.8085 & 0.5000 & 0.9697 & 0.9920 & 0.8840 \\
			\bottomrule
		\end{tabular}
	\end{minipage}\hfill
	\begin{minipage}[t]{0.49\textwidth}
		\centering
		\scriptsize
		\setlength{\tabcolsep}{1pt}
		\caption{Classification results of \(F_2\)-CEQRP-ML (\(\mathcal{OD}_5\)).}
		\label{tab30}
		\begin{tabular}{lccccccc}
			\toprule
			Models & LR & SVM & ANN & GB & AdaBoost & EM & BRLS \\
			\midrule
			CV           & 0.8967 & 0.8864 & 0.8866 & 0.8848 & 0.8694 & 0.9075 & 0.8980 \\
			Validation   & 0.9069 & 0.9025 & 0.9026 & 0.8969 & 0.8750 & 0.9075 & 0.9131 \\
			WernerTest   & 0.5000 & 0.6328 & 0.5000 & 0.8723 & 0.8032 & 0.5000 & 0.8695 \\
			IsotropicTest& 0.5000 & 0.6065 & 0.5000 & 0.7976 & 0.9787 & 0.8502 & 0.8893 \\
			\bottomrule
		\end{tabular}
	\end{minipage}
\end{table}

% ================= 第 5 组：tab31 + tab32 =================
\begin{table}[!htbp]
	\centering
	\begin{minipage}[t]{0.49\textwidth}
		\centering
		\scriptsize
		\setlength{\tabcolsep}{1pt}
		\caption{Classification results of \(F_2\)-CEQRP-ML (\(\mathcal{OD}_6\)).}
		\label{tab31}
		\begin{tabular}{lccccccc}
			\toprule
			Models & LR & SVM & ANN & GB & AdaBoost & EM & BRLS \\
			\midrule
			CV           & 0.8891 & 0.8802 & 0.8811 & 0.8789 & 0.8633 & 0.8823 & 0.8834 \\
			Validation   & 0.8988 & 0.8875 & 0.8925 & 0.8919 & 0.8838 & 0.8956 & 0.8769 \\
			WernerTest   & 0.5000 & 0.5000 & 0.5000 & 0.5000 & 0.8210 & 0.5000 & 0.9400 \\
			IsotropicTest& 0.5000 & 0.5000 & 0.8750 & 0.5000 & 0.9715 & 0.8450 & 0.8575 \\
			\bottomrule
		\end{tabular}
	\end{minipage}\hfill
	\begin{minipage}[t]{0.49\textwidth}
		\centering
		\scriptsize
		\setlength{\tabcolsep}{1pt}
		\caption{Classification results of \(F_2\)-CEQRP-ML (\(\mathcal{OD}_7\)).}
		\label{tab32}
		\begin{tabular}{lccccccc}
			\toprule
			Models & LR & SVM & ANN & GB & AdaBoost & EM & BRLS \\
			\midrule
			CV           & 0.8834 & 0.9175 & 0.9196 & 0.9314 & 0.8731 & 0.9395 & 0.8911 \\
			Validation   & 0.8965 & 0.9295 & 0.9270 & 0.9380 & 0.8765 & 0.9515 & 0.9000 \\
			WernerTest   & 0.5000 & 0.5000 & 0.5000 & 0.5000 & 0.8117 & 0.5000 & 0.9933 \\
			IsotropicTest& 0.5000 & 0.5000 & 0.8300 & 0.5000 & 0.9752 & 0.5000 & 0.9025 \\
			\bottomrule
		\end{tabular}
	\end{minipage}
\end{table}

	% =================qrp 第 1  组：tab31 + tab32 =================

\begin{table}[!htbp]
	\centering
	\begin{minipage}[t]{0.49\textwidth}
		\centering
		\scriptsize
		\setlength{\tabcolsep}{1pt}
		\caption{Classification results of \(F_1\)-QRP-ML (\(\mathcal{OD}_3\)).}
		\label{tab40}
		\begin{tabular}{lccccccc}
			\toprule
			Models & LR & SVM & ANN & GB & AdaBoost & EM & BRLS \\
			\midrule
			CV           & 0.8863 & 0.8756 & 0.8739 & 0.8719 & 0.8430 & 0.8763  & 0.7424 \\
			Validation   & 0.8675 & 0.8575 & 0.8619 & 0.8519 & 0.8206 & 0.8625 & 0.7516 \\
			WernerTest   & 0.6385 & 0.5000 & 0.5000 & 0.6503 & 0.5000 & 0.5000 & 0.5000 \\
			IsotropicTest& 0.8485 & 0.5000 & 0.8258 & 0.8905 & 0.8690 & 0.5000 & 0.5000 \\
			\bottomrule
		\end{tabular}
	\end{minipage}\hfill
	\begin{minipage}[t]{0.49\textwidth}
		\centering
		\scriptsize
		\setlength{\tabcolsep}{1pt}
		\caption{Classification results of \(F_1\)-QRP-ML (\(\mathcal{OD}_4\)).}
		\label{tab41}
		\begin{tabular}{lccccccc}
			\toprule
			Models & LR & SVM & ANN & GB & AdaBoost & EM & BRLS \\
			\midrule
			CV           & 0.7713 & 0.7414 & 0.7498 & 0.7633 & 0.7333 & 0.7652 & 0.7011 \\
			Validation   & 0.7756 & 0.7575 & 0.7575 & 0.7713 & 0.7469 & 0.7744 & 0.7032 \\
			WernerTest   & 0.6048 & 0.5000 & 0.5000 & 0.6340 & 0.5000 & 0.5000 & 0.5000 \\
			IsotropicTest& 0.7725 & 0.5000 & 0.8133 & 0.7088 & 0.8643 & 0.6305 & 0.7723 \\
			\bottomrule
		\end{tabular}
	\end{minipage}
\end{table}

% =================QRP 第 2 组：tab25 + tab26 =================
\begin{table}[!htbp]
	\centering
	\begin{minipage}[t]{0.49\textwidth}
		\centering
		\scriptsize
		\setlength{\tabcolsep}{1pt}
		\caption{Classification results of \(F_1\)-QRP-ML (\(\mathcal{OD}_5\)).}
		\label{tab42}
		\begin{tabular}{lccccccc}
			\toprule
			Models & LR & SVM & ANN & GB & AdaBoost & EM & BRLS \\
			\midrule
			CV           & 0.8577 & 0.8477 & 0.8420 & 0.8436 & 0.8063 & 0.8525 & 0.7346 \\
			Validation   & 0.8694 & 0.8525 & 0.8488 & 0.8444 & 0.8106 & 0.8544 & 0.7950 \\
			WernerTest   & 0.6303 & 0.5000 & 0.5000 & 0.5000 & 0.5000 & 0.5000 & 0.5000 \\
			IsotropicTest& 0.9583 & 0.5000 & 0.8755 & 0.8758 & 0.8700 & 0.5000 & 0.5000 \\
			\bottomrule
		\end{tabular}
	\end{minipage}\hfill
	\begin{minipage}[t]{0.49\textwidth}
		\centering
		\scriptsize
		\setlength{\tabcolsep}{1pt}
		\caption{Classification results of \(F_1\)-QRP-ML (\(\mathcal{OD}_6\)).}
		\label{tab43}
		\begin{tabular}{lccccccc}
			\toprule
			Models & LR & SVM & ANN & GB & AdaBoost & EM & BRLS \\
			\midrule
			CV           & 0.8527 & 0.8422 & 0.8380 & 0.8370 & 0.8011 & 0.8489 & 0.7261 \\
			Validation   & 0.8481 & 0.8488 & 0.8406 & 0.8388 & 0.8150 & 0.8563 & 0.7348 \\
			WernerTest   & 0.6435 & 0.5000 & 0.6008 & 0.5000 & 0.5000 & 0.5000 & 0.5000 \\
			IsotropicTest& 0.8465 & 0.5000 & 0.6060 & 0.6780 & 0.8713 & 0.5000 & 0.9512 \\
			\bottomrule
		\end{tabular}
	\end{minipage}
\end{table}

% =================qrp 第 3 组：tab27 + tab28 =================
\begin{table}[!htbp]
	\centering
	\begin{minipage}[t]{0.49\textwidth}
		\centering
		\scriptsize
		\setlength{\tabcolsep}{1pt}
		\caption{Classification results of \(F_1\)-QRP-ML (\(\mathcal{OD}_7\)).}
		\label{tab44}
		\begin{tabular}{lccccccc}
			\toprule
			Models & LR & SVM & ANN & GB & AdaBoost & EM & BRLS \\
			\midrule
			CV           & 0.8522 & 0.8906 & 0.9039 & 0.8598 & 0.7947 & 0.9147 & 0.7132 \\
			Validation   & 0.8625 & 0.9169 & 0.9219 & 0.8681 & 0.7825 & 0.9406 & 0.8219 \\
			WernerTest   & 0.6370 & 0.5000 & 0.6228 & 0.6055 & 0.5000 & 0.5000 & 0.5000 \\
			IsotropicTest& 0.9935 & 0.5000 & 0.6943 & 0.8838 & 0.8723 & 0.5000 & 0.9120 \\
			\bottomrule
		\end{tabular}
	\end{minipage}\hfill
	\begin{minipage}[t]{0.49\textwidth}
		\centering
		\scriptsize
		\setlength{\tabcolsep}{1pt}
		\caption{Classification results of \(F_2\)-QRP-ML (\(\mathcal{OD}_3\)).}
		\label{tab45}
		\begin{tabular}{lccccccc}
			\toprule
			Models & LR & SVM & ANN & GB & AdaBoost & EM & BRLS \\
			\midrule
			CV           & 0.9213 & 0.9149 & 0.9170 & 0.9163 & 0.8929 & 0.9201 & 0.9174 \\
			Validation   & 0.9200 & 0.9143 & 0.9193 & 0.9129 & 0.8886 & 0.9161 & 0.9261 \\
			WernerTest   & 0.5000 & 0.5000 & 0.5000 & 0.5000 & 0.5000 & 0.5000 & 0.5000 \\
			IsotropicTest& 0.5000 & 0.5000 & 0.5000 & 0.5000 & 0.5000 & 0.7949 & 0.7453 \\
			\bottomrule
		\end{tabular}
	\end{minipage}
\end{table}

% ================= F2 第 4 组：tab29 + tab30 =================
\begin{table}[!htbp]
	\centering
	\begin{minipage}[t]{0.49\textwidth}
		\centering
		\scriptsize
		\setlength{\tabcolsep}{1pt}
		\caption{Classification results of \(F_2\)-QRP-ML (\(\mathcal{OD}_4\)).}
		\label{tab46}
		\begin{tabular}{lccccccc}
			\toprule
			Models & LR & SVM & ANN & GB & AdaBoost & EM & BRLS \\
			\midrule
			CV           & 0.8013 & 0.7917 & 0.7970 & 0.8030 & 0.7703 & 0.8697 & 0.8456 \\
			Validation   & 0.7969 & 0.8038 & 0.8075 & 0.8038 & 0.7875 & 0.8675 & 0.8532 \\
			WernerTest   & 0.5000 & 0.5000 & 0.5000 & 0.6588 & 0.5000 & 0.6309 & 0.5000 \\
			IsotropicTest& 0.5000 & 0.5000 & 0.5000 & 0.7870 & 0.5000 & 0.9920 & 0.8932 \\
			\bottomrule
		\end{tabular}
	\end{minipage}\hfill
	\begin{minipage}[t]{0.49\textwidth}
		\centering
		\scriptsize
		\setlength{\tabcolsep}{1pt}
		\caption{Classification results of \(F_2\)-QRP-ML (\(\mathcal{OD}_5\)).}
		\label{tab47}
		\begin{tabular}{lccccccc}
			\toprule
			Models & LR & SVM & ANN & GB & AdaBoost & EM & BRLS \\
			\midrule
			CV           & 0.8978 & 0.8902 & 0.8914 & 0.8881 & 0.8722 & 0.9075 & 0.8980 \\
			Validation   & 0.9044 & 0.9063 & 0.8938 & 0.9013 & 0.8719 & 0.9075 & 0.9131 \\
			WernerTest   & 0.5000 & 0.5000 & 0.5000 & 0.5000 & 0.5000 & 0.5000 & 0.5000 \\
			IsotropicTest& 0.5000 & 0.5000 & 0.5000 & 0.5000 & 0.5000 & 0.8502 & 0.7469 \\
			\bottomrule
		\end{tabular}
	\end{minipage}
\end{table}

% =================F2 第 5 组：tab31 + tab32 =================
\begin{table}[!htbp]
	\centering
	\begin{minipage}[t]{0.49\textwidth}
		\centering
		\scriptsize
		\setlength{\tabcolsep}{1pt}
		\caption{Classification results of \(F_2\)-QRP-ML (\(\mathcal{OD}_6\)).}
		\label{tab48}
		\begin{tabular}{lccccccc}
			\toprule
			Models & LR & SVM & ANN & GB & AdaBoost & EM & BRLS \\
			\midrule
			CV           & 0.8902 & 0.8784 & 0.8808 & 0.8863 & 0.8636 & 0.8823 & 0.8753 \\
			Validation   & 0.9019 & 0.8856 & 0.8931 & 0.8844 & 0.8575 & 0.8956 & 0.8816 \\
			WernerTest   & 0.5000 & 0.5000 & 0.5000 & 0.5000 & 0.5000 & 0.5000 & 0.5000 \\
			IsotropicTest& 0.5000 & 0.5000 & 0.5000 & 0.5000 & 0.5000 & 0.8450 & 0.7763 \\
			\bottomrule
		\end{tabular}
	\end{minipage}\hfill
	\begin{minipage}[t]{0.49\textwidth}
		\centering
		\scriptsize
		\setlength{\tabcolsep}{1pt}
		\caption{Classification results of \(F_2\)-QRP-ML (\(\mathcal{OD}_7\)).}
		\label{tab49}
		\begin{tabular}{lccccccc}
			\toprule
			Models & LR & SVM & ANN & GB & AdaBoost & EM & BRLS \\
			\midrule
			CV           & 0.8920 & 0.9078 & 0.9161 & 0.9061 & 0.8622 & 0.9395 & 0.8814 \\
			Validation   & 0.8838 & 0.9113 & 0.9188 & 0.8956 & 0.8606 & 0.9515 & 0.8903 \\
			WernerTest   & 0.5000 & 0.5000 & 0.5000 & 0.5000 & 0.5000 & 0.5000 & 0.5000 \\
			IsotropicTest& 0.5000 & 0.5000 & 0.5000 & 0.5000 & 0.5000 & 0.5000 & 0.5000 \\
			\bottomrule
		\end{tabular}
	\end{minipage}
\end{table}

\newpage
\subsection{Theorems and proofs}
\label{app:thm_proof}

This section contains the proofs of Theorems~1 and~2 of the main text.

We consider mixed two-qutrit states for which a projective measurement on Alice's subsystem produces three linearly independent pure conditional states on Bob's subsystem. Up to a local unitary transformation on Alice's subsystem, such states can be expressed as
\begin{equation}
	\begin{aligned}
		\rho_{AB}
		={}&
		\sum_{j=0}^{2}\mu_{j+1}
		\ket{j}\!\bra{j}\otimes
		\ket{\phi_{j+1}}\!\bra{\phi_{j+1}}
		+ \ket{0}\!\bra{1}\otimes\mathcal{A}
		+ \ket{1}\!\bra{0}\otimes\mathcal{A}^{\dagger}
		+\ket{0}\!\bra{2}\otimes\mathcal{B}
		+
		\ket{2}\!\bra{0}\otimes\mathcal{B}^{\dagger}
		+
		\ket{1}\!\bra{2}\otimes\mathcal{C}
		+
		\ket{2}\!\bra{1}\otimes\mathcal{C}^{\dagger},
	\end{aligned}
	\label{eq:conditional_pure_state_decomposition}
\end{equation}
where \(\mu_j\geq0\), \(\sum_{j=1}^{3}\mu_j=1\), and \(\{\ket{\phi_j}\}_{j=1}^{3}\) are linearly independent normalized vectors in \(\mathbb C^3\). The operators \(\mathcal A\), \(\mathcal B\), and \(\mathcal C\) represent the off-diagonal coherence blocks.

\noindent\textbf{Theorem 1(AVN criterion).}
\textit{A two-qutrit state of the above form
	is unsteerable  from Alice to Bob under any projective measurement assemblage if and only if
	$\mathcal A=\mathcal B=\mathcal C=0$.}

\begin{proof}
	Suppose that \(\rho_{AB}\) is unsteerable from Alice to Bob under any projective measurement assemblage. Then it  is unsteerable  from Alice to Bob under
 the  projective measurement assemblage $\mathcal{M}_A=\{M^{x}: x=0,1,2\ldots,6\}$,
where  $M^{x}=\left\{M_{a|x}\equiv|e_{ax}\>\<e_{ax}|\right\}_{a=0}^2$ with
	\begin{align*}
&\{|e_{01}\>=|0\>, |e_{11}\>=|1\>,|e_{21}\>=|2\>\},\\
		&\{|e_{02}\>=(\ket{0}+\ket{1})/\sqrt{2},|e_{12}\>=(\ket{0}-\ket{1})/\sqrt{2},|e_{22}\>=\ket{2}\},\\
		&\{|e_{03}\>=(\ket{0}+{\rm i}\ket{1})/\sqrt{2},|e_{13}\>=(\ket{0}-{\rm i}\ket{1})/\sqrt{2},|e_{23}\>=\ket{2}\},\\
		&\{|e_{04}\>=(\ket{0}+\ket{2})/\sqrt{2},|e_{14}\>=(\ket{0}-\ket{2})/\sqrt{2},|e_{24}\>=\ket{1}\},\\
		&\{|e_{05}\>=(\ket{0}+{\rm i}\ket{2})/\sqrt{2},|e_{15}\>=(\ket{0}-{\rm i}\ket{2})/\sqrt{2},|e_{25}\>=\ket{1}\},\\
		&\{|e_{06}\>=(\ket{1}+\ket{2})/\sqrt{2},|e_{15}\>=(\ket{1}-\ket{2})/\sqrt{2},|e_{25}\>=\ket{0}\},\\
		&\{|e_{07}\>=(\ket{1}+{\rm i}\ket{2})/\sqrt{2},|e_{17}\>=(\ket{1}-{\rm i}\ket{2})/\sqrt{2},|e_{27}\>=\ket{0}\}.
	\end{align*}
Equivalently, the assemble $\sigma_{a|x}^B=\operatorname{Tr}_A[(M_{a|x}\ot I_B)\rho_{AB}]$ admits an LHS model:
\begin{equation}\label{T-Proof-LHSM}
\sigma_{a|x}^B=\sum_{\lm=1}^N\pi_\lm P_A(a|x,\lm)\sigma_\lm,\ \ \forall x=0,1,2,\ldots,6, a=0,1,2.
\end{equation}
where $\pi_\lm$ and $P_A(a|x,\lm)$ are probability distributions of $\lm$ and $a$, respectively, and $\sigma_\lm
$ are states of Bob's system. We may assume that $\pi_\lm>0$ for all $\lm=0,1,\ldots,6$.

Using \eqref{T-Proof-LHSM} for $x=0$ and $a=j$ yields that
$$\mu_{j+1}\ket{\phi_{j+1}}\!\bra{\phi_{j+1}}=\sum_{\lm=1}^N\pi_\lm P_A(j|0,\lm)\sigma_\lm,$$
i.e.,
$$\ket{\phi_{j+1}}\!\bra{\phi_{j+1}}=\sum_{\lm=1}^N\pi_\lm \mu_{j+1}^{-1}P_A(j|0,\lm)\sigma_\lm.$$
Since $\sum_{\lm=1}^N\pi_\lm \mu_{j+1}^{-1}P_A(j|0,\lm)=1$ and $\ket{\phi_{j+1}}\!\bra{\phi_{j+1}}$ is an extreme point of convex set $\mathcal{D}(\C^3)$, we see  that for each $\lm\in[N]$, either  $\sigma_\lm=\ket{\phi_{j+1}}\!\bra{\phi_{j+1}}$ or  $P_A(j|0,\lm)=0$. Since $\sum_{j=0}^2P_A(j|0,\lm)=1$ for all $\lm\in[N]$, we see for each $\lm\in[N]$, there exists a $j\in\{0,1,2\}$ such that $P_A(j|0,\lm)>0$ and so  $\sigma_\lm=\ket{\phi_{j+1}}\!\bra{\phi_{j+1}}$. This shows that
each  hidden state  $\sigma_\lm$ is one of
	\(
	\ket{\phi_1}\!\bra{\phi_1},
	\ket{\phi_2}\!\bra{\phi_2},
	\ket{\phi_3}\!\bra{\phi_3}.
	\)
Put
$$S_j=\{\lm\in[N]:\, \sigma_\lm=\ket{\phi_{j+1}}\!\bra{\phi_{j+1}}\}(j=0,1,2),$$
then we get a partition $\{S_0,S_1,S_2\}$ of $[N]$ so that the LHS model (\ref{T-Proof-LHSM}) becomes
\begin{eqnarray}\label{T-Proof-LHSM-3}
\sigma_{a|x}^B&=&\sum_{\lm\in S_0}\pi_\lm P_A(a|x,\lm)\cdot\ket{\phi_1}\!\bra{\phi_1}
+\sum_{\lm\in S_1}\pi_\lm P_A(a|x,\lm)\cdot\ket{\phi_2}\!\bra{\phi_2}
+\sum_{\lm\in S_2}\pi_\lm P_A(a|x,\lm)\cdot\ket{\phi_3}\!\bra{\phi_3}\nonumber\\
&=:&f_1(a,x)\ket{\phi_1}\!\bra{\phi_1}
+f_2(a,x)\ket{\phi_2}\!\bra{\phi_2}
+f_3(a,x)\ket{\phi_3}\!\bra{\phi_3}
\end{eqnarray}
for all $x=0,1,2,\ldots,6,$ and $a=0,1,2,$ here when $S_j=\emptyset$, the corresponding sum equals to $0$.

Direct evaluation of the assemblage elements for the six remaining measurements yields
\begingroup
\setlength{\arraycolsep}{1pt}
\begin{align*}
	\sigma_{0|1}^B
	&=\frac12\left(\mu_1|\phi_1\rangle\langle\phi_1|+\mu_2|\phi_2\rangle\langle\phi_2|+\mathcal A+\mathcal A^\dagger\right),
	&
	\sigma_{1|1}^B
	&=\frac12\left(\mu_1|\phi_1\rangle\langle\phi_1|+\mu_2|\phi_2\rangle\langle\phi_2|-\mathcal A-\mathcal A^\dagger\right),
	&
	\sigma_{2|1}^B&=\mu_3|\phi_3\rangle\langle\phi_3|,\\
	%\label{eq:thm1_assemblage_Areal} \\
	\sigma_{0|2}^B
	&=\frac12\left(\mu_1|\phi_1\rangle\langle\phi_1|+\mu_2|\phi_2\rangle\langle\phi_2|+\mathrm i\mathcal A-\mathrm i\mathcal A^\dagger\right),
	&
	\sigma_{1|2}^B
	&=\frac12\left(\mu_1|\phi_1\rangle\langle\phi_1|+\mu_2|\phi_2\rangle\langle\phi_2|-\mathrm i\mathcal A+\mathrm i\mathcal A^\dagger\right),
	&
	\sigma_{2|2}^B&=\mu_3|\phi_3\rangle\langle\phi_3|,\\
	%\label{eq:thm1_assemblage_Aimag}\\
	\sigma_{0|3}^B
	&=\frac12\left(\mu_1|\phi_1\rangle\langle\phi_1|+\mu_3|\phi_3\rangle\langle\phi_3|+\mathcal B+\mathcal B^\dagger\right),
	&
	\sigma_{1|3}^B
	&=\frac12\left(\mu_1|\phi_1\rangle\langle\phi_1|+\mu_3|\phi_3\rangle\langle\phi_3|-\mathcal B-\mathcal B^\dagger\right),
	&
	\sigma_{2|3}^B&=\mu_2|\phi_2\rangle\langle\phi_2|,\\
	\sigma_{0|4}^B
	&=\frac12\left(\mu_1|\phi_1\rangle\langle\phi_1|+\mu_3|\phi_3\rangle\langle\phi_3|+\mathrm i\mathcal B-\mathrm i\mathcal B^\dagger\right),
	&
	\sigma_{1|4}^B
	&=\frac12\left(\mu_1|\phi_1\rangle\langle\phi_1|+\mu_3|\phi_3\rangle\langle\phi_3|-\mathrm i\mathcal B+\mathrm i\mathcal B^\dagger\right),
	&
	\sigma_{2|4}^B&=\mu_2|\phi_2\rangle\langle\phi_2|,\\
	\sigma_{0|5}^B
	&=\frac12\left(\mu_2|\phi_2\rangle\langle\phi_2|+\mu_3|\phi_3\rangle\langle\phi_3|+\mathcal C+\mathcal C^\dagger\right),
	&
	\sigma_{1|5}^B
	&=\frac12\left(\mu_2|\phi_2\rangle\langle\phi_2|+\mu_3|\phi_3\rangle\langle\phi_3|-\mathcal C-\mathcal C^\dagger\right),
	&
	\sigma_{2|5}^B&=\mu_1|\phi_1\rangle\langle\phi_1|,\\
	\sigma_{0|6}^B
	&=\frac12\left(\mu_2|\phi_2\rangle\langle\phi_2|+\mu_3|\phi_3\rangle\langle\phi_3|+\mathrm i\mathcal C-\mathrm i\mathcal C^\dagger\right),
	&
	\sigma_{1|6}^B
	&=\frac12\left(\mu_2|\phi_2\rangle\langle\phi_2|+\mu_3|\phi_3\rangle\langle\phi_3|-\mathrm i\mathcal C+\mathrm i\mathcal C^\dagger\right),
	&
	\sigma_{2|6}^B&=\mu_1|\phi_1\rangle\langle\phi_1|.
	%\label{eq:thm1_remaining_assemblages}
\end{align*}
\endgroup

Taking suitable differences of these assemblage elements gives
\[
\begin{aligned}
	\mathcal A
	&=\frac12\left[
	(\sigma_{0|1}^B-\sigma_{1|1}^B)
	-\mathrm i(\sigma_{0|2}^B-\sigma_{1|2}^B)
	\right],\\
	\mathcal B
	&=\frac12\left[
	(\sigma_{0|3}^B-\sigma_{1|3}^B)
	-\mathrm i(\sigma_{0|4}^B-\sigma_{1|4}^B)
	\right],\\
	\mathcal C
	&=\frac12\left[
	(\sigma_{0|5}^B-\sigma_{1|5}^B)
	-\mathrm i(\sigma_{0|6}^B-\sigma_{1|6}^B)
	\right].
\end{aligned}
\]
Substituting Eq.~\eqref{T-Proof-LHSM-3} into the above expressions yields
\begin{align}
	\mathcal A
	&=\alpha_1|\phi_1\rangle\langle\phi_1|+\beta_1|\phi_2\rangle\langle\phi_2|+\gamma_1|\phi_3\rangle\langle\phi_3|,
	\nonumber\\
	\mathcal B
	&=\alpha_2|\phi_1\rangle\langle\phi_1|+\beta_2|\phi_2\rangle\langle\phi_2|+\gamma_2|\phi_3\rangle\langle\phi_3|,
	\nonumber\\
	\mathcal C
	&=\alpha_3|\phi_1\rangle\langle\phi_1|+\beta_3|\phi_2\rangle\langle\phi_2|+\gamma_3|\phi_3\rangle\langle\phi_3|,
	\label{eq:thm1_coherence_expansion}
\end{align}
where, for \(r=1,2,3\), the coefficients are defined directly by \(f_r(a,x)\) as
\begin{align*}
	\alpha_r
	&=\frac12\left[f_1(0,2r-1)-f_1(1,2r-1)-\mathrm{i}\left(f_1(0,2r)-f_1(1,2r)\right)\right],\\
	\beta_r
	&=\frac12\left[f_2(0,2r-1)-f_2(1,2r-1)-\mathrm{i}\left(f_2(0,2r)-f_2(1,2r)\right)\right],\\
	\gamma_r
	&=\frac12\left[f_3(0,2r-1)-f_3(1,2r-1)-\mathrm{i}\left(f_3(0,2r)-f_3(1,2r)\right)\right].
\end{align*}

Substituting Eq.~\eqref{eq:thm1_coherence_expansion} into the expression \eqref{eq:conditional_pure_state_decomposition} of \(\rho_{AB}\) and collecting the terms multiplying \(|\phi_j\rangle\langle\phi_j|\) gives
\begin{equation}\label{T1-rho-1}
\rho_{AB}=T_1\otimes |\phi_1\rangle\langle\phi_1|+T_2\otimes |\phi_2\rangle\langle\phi_2|+T_3\otimes |\phi_3\rangle\langle\phi_3|,
\end{equation}
where
\[
T_1=
\begin{pmatrix}
	\mu_1 & \alpha_1 & \alpha_2\\
	\overline{\alpha}_1 & 0 & \alpha_3\\
	\overline{\alpha}_2 & \overline{\alpha}_3 & 0
\end{pmatrix},
\quad
T_2=
\begin{pmatrix}
	0 & \beta_1 & \beta_2\\
	\overline{\beta}_1 & \mu_2 & \beta_3\\
	\overline{\beta}_2 & \overline{\beta}_3 & 0
\end{pmatrix},
\quad
T_3=
\begin{pmatrix}
	0 & \gamma_1 & \gamma_2\\
	\overline{\gamma}_1 & 0 & \gamma_3\\
	\overline{\gamma}_2 & \overline{\gamma}_3 & \mu_3
\end{pmatrix}.
\]

We first show that \(T_1\ge0\). Suppose, to the contrary, that \(T_1\) has a negative eigenvalue \(\upsilon_1<0\) with normalized eigenvector \(|\chi_1\rangle\).
Since \(|\phi_1\rangle,|\phi_2\rangle,|\phi_3\rangle\) are linearly independent, $|\phi_1\rangle\notin\mathrm{span}\{|\phi_2\>,|\phi_2\>\}=\{|\phi_2\>,|\phi_2\>\}^{\bot\bot}.$ Thus,
there exists a normalized vector \(|\phi'_1\rangle \in \{|\phi_2\>,|\phi_2\>\}^{\bot}\)  such that \(\langle\phi'_1|\phi_1\rangle\neq0\). The operator
\(Q_1=|\chi_1\rangle\langle\chi_1|\otimes|\phi'_1\rangle\langle\phi'_1|\)
is a projection. Therefore, since \(\rho_{AB}\ge0\), \(\operatorname{Tr}(Q_1\rho_{AB})\geq0\). However, Eq.~\eqref{T1-rho-1} implies
\[
\operatorname{Tr}(Q_1\rho_{AB})=\upsilon_1|\langle\phi'_1|\phi_1\rangle|^2<0,
\]
a contradiction. Hence \(T_1\ge0\) and so \(\alpha_1=\alpha_2=\alpha_3=0\). The same argument shows that \(T_2,T_3\ge0\), and consequently \(\beta_j=\gamma_j(j=1,2,3)\).
It follows from Eq.~\eqref{eq:thm1_coherence_expansion} that \(\mathcal A=\mathcal B=\mathcal C=0\).

This proves the necessity.

Conversely, if \(\mathcal A=\mathcal B=\mathcal C=0\), then
\[
\rho_{AB}
=
\sum_{j=0}^2
\mu_{j+1}\,|j\rangle\langle j|\otimes |\phi_{j+1}\rangle\langle\phi_{j+1}|
\]
is separable, and therefore unsteerable from Alice to Bob under any projective measurement assemblage. This proves sufficiency and completes the proof.
\end{proof}

For a family of unnormalized conditional states
\[
\sigma_t=
\begin{pmatrix}
	a_t & b_t & c_t\\
	\overline{b_t} & d_t & e_t\\
	\overline{c_t} & \overline{e_t} & f_t
\end{pmatrix},
\]
define the transverse coherence and the longitudinal population leakage as
\begin{equation}
	T_t=\sqrt{3\bigl(|b_t|^2+|c_t|^2\bigr)},
	\qquad
	D_t=d_t+f_t.
	\label{eq:def_TD}
\end{equation}

\noindent\textbf{Lemma 1.}
Let \(\rho_{AB}\) be a state on \(\mathcal H_A\otimes\mathcal H_B\) and
let $\mathcal{M}_A=\left\{\{M_{a|x}\}_{a=1}^{o_A}:x\in[m_A]\right\}$
be a measurement assemblage of subsystem $A$ and put
$$\rho^B_{a|x}={\rm{Tr}}_A\left[(M_{a|x}\ot I_B)\rho_{AB}\right],$$
denoting the assemblage of unnormalized states  on Bob's side. Assume that the assemblage $\{\rho^B_{a|x}\}_{a\in[o_A],x\in[m_A]}$ admits an integral LHS model
\begin{equation}\label{C-LHSM}
\rho^B_{a|x}=\int_{\Omega}P_A(a|x,\lambda)\,\sigma(\lambda)\,d\mu(\lambda),\  \forall x,a,
\end{equation}
where $(\Omega,\Sigma,\mu)$ is a probability space,  $\sigma:\Omega\to \D(\H_B)$ is a measurable function,
$P_A(a|x,\lambda)$ is measurable in $\lambda$ and forms a probability distribution over outcomes $a$ for each fixed $x,\lambda$. Then it admits finite-sum LHS model:
\begin{equation}\label{UD-LHSM}
\rho^B_{a|x}=\sum_{\lm'\in \Lambda_D}q_{\lambda'}D(a|x,\lm')\widetilde\rho_{\lambda'},\,\  \forall x,a,
\end{equation}
where $\Lambda_D$ denotes the set of all maps $\lm'$ from $[m_A]$ to $[o_A]$, \(\widetilde\rho_{\lambda'}\) are states of Bob's system,  $q_{\lambda'}$ and $D(a|x,\lm')=\delta_{a,\lm'(x)}$ are probability distributions of $\lm'$ and $a$, respectively.

\begin{proof}
	Let \(\Lambda_D=\{\lambda':[m_A]\to[o_A]\}\)
	be the set of all deterministic strategies, and define \(D(a|x,\lambda')=\delta_{a,\lambda'(x)}.\)
Thus, for each fixed \(x\) and \(\lambda'\),
	\(\{D(a|x,\lambda')\}_{a=1}^{o_A}\) is a probability distribution.
	
	For each \(\lambda\in\Omega\) and \(\lambda'\in\Lambda_D\), define
	\[
	p(\lambda'|\lambda)
	=
	\prod_{x=1}^{m_A}P_A(\lambda'(x)|x,\lambda).
	\]
	Since \(P_A(a|x,\lambda)\) is measurable in \(\lambda\),
	\(p(\lambda'|\lambda)\) is measurable in \(\lambda\). Moreover,
	\[
	\begin{aligned}
		\sum_{\lambda'\in\Lambda_D}p(\lambda'|\lambda)
		&=
		\sum_{a_1,\ldots,a_{m_A}=1}^{o_A}
		\prod_{x=1}^{m_A}P_A(a_x|x,\lambda)\\
		&=
		\prod_{x=1}^{m_A}
		\left(\sum_{a_x=1}^{o_A}P_A(a_x|x,\lambda)\right)
		=1.
	\end{aligned}
	\]
	For each \(x\in[m_A]\) and \(a\in[o_A]\), we also have
	\[
	\begin{aligned}
		\sum_{\lambda'\in\Lambda_D}
		p(\lambda'|\lambda)D(a|x,\lambda')
		&=
		\sum_{\lambda':\,\lambda'(x)=a}
		\prod_{y=1}^{m_A}P_A(\lambda'(y)|y,\lambda)\\
		&=
		P_A(a|x,\lambda)
		\prod_{y\ne x}
		\left(\sum_{b=1}^{o_A}P_A(b|y,\lambda)\right)\\
		&=P_A(a|x,\lambda).
	\end{aligned}
	\]
	Hence,	
	\begin{equation}\label{PD}
	P_A(a|x,\lambda)
	=\sum_{\lambda'\in\Lambda_D}
	p(\lambda'|\lambda)D(a|x,\lambda'), \, \forall x,a,\lambda.
	\end{equation}	
Finding sums of two sides over $a\in[o_A]$  implies that
\begin{equation}\label{OOO}
\sum_{\lambda'\in\Lambda_D}p(\lambda'|\lambda)=1,\ \forall \lm\in \Lambda_D.\end{equation}

	For each \(\lambda'\in\Lambda_D\), define the positive operator
	\[
	\sigma_{\lambda'}
	=
	\int_\Omega
	p(\lambda'|\lambda)\sigma(\lambda)\,d\mu(\lambda).
	\]
	The integral is well defined: its integrand is measurable and
	\[
	\|p(\lambda'|\lambda)\sigma(\lambda)\|_1
	=
	p(\lambda'|\lambda)\leq 1.
	\]
	Substituting the deterministic-response decomposition \eqref{PD} into
	\eqref{C-LHSM} and interchanging the finite sum with the integral gives
	\begin{eqnarray}
		\rho^B_{a|x}
		&=&
		\int_\Omega
		\left(
		\sum_{\lambda'\in\Lambda_D}
		p(\lambda'|\lambda)D(a|x,\lambda')
		\right)
		\sigma(\lambda)\,d\mu(\lambda)\\
		&=&
		\sum_{\lambda'\in\Lambda_D}
		D(a|x,\lambda')\sigma_{\lambda'},
		\qquad \forall x,a. \label{SDD}
	\end{eqnarray}
		Furthermore, we get from Eqs. \eqref{OOO} and \eqref{C-LHSM} that
\[		\begin{aligned}
	\sum_{\lambda'\in\Lambda_D}\sigma_{\lambda'}
		&=
		\int_\Omega
		\left(\sum_{\lambda'\in\Lambda_D}p(\lambda'|\lambda)\right)
		\sigma(\lambda)\,d\mu(\lambda)\\
		&=
		\int_\Omega \sigma(\lambda)\,d\mu(\lambda)\\
&=\sum_{a=1}^{o_A}\rho^B_{a|x}\\
&=\operatorname{Tr}_A\rho_{AB}\\
		&=\rho_B.%\label{SDD}
	\end{aligned}
	\]
	
	For each \(\lambda'\in\Lambda_D\), define \(
	q_{\lambda'}=\operatorname{Tr}(\sigma_{\lambda'}).
	\) 	Since \(\sigma_{\lambda'}\geq0\), we have
	\(q_{\lambda'}\geq0\). Moreover,
	\[
	\sum_{\lambda'\in\Lambda_D}q_{\lambda'}
	=
	\operatorname{Tr}\left(
	\sum_{\lambda'\in\Lambda_D}\sigma_{\lambda'}
	\right)
	=
	\operatorname{Tr}(\rho_B)
	=
	1.
	\]
	Thus, \(\{q_{\lambda'}\}_{\lambda'\in\Lambda_D}\)
	is a probability distribution on \(\Lambda_D\).
	
	For \(q_{\lambda'}>0\), define
	$	\widetilde\rho_{\lambda'}
	=
	\frac{\sigma_{\lambda'}}{q_{\lambda'}}.
	$
	Then
	\[
	\widetilde\rho_{\lambda'}\ge 0,
	\qquad
	\operatorname{Tr}(\widetilde\rho_{\lambda'})
	=
	\frac{\operatorname{Tr}(\sigma_{\lambda'})}{q_{\lambda'}}
	=
	1.
	\]
	For \(q_{\lambda'}=0\), the positivity of \(\sigma_{\lambda'}\) and
	\(\operatorname{Tr}(\sigma_{\lambda'})=0\) imply that
	\(\sigma_{\lambda'}=0\). In this case, choose an arbitrary state
	\(\omega_B\in\mathcal D(\mathcal H_B)\) and set
	\(\widetilde\rho_{\lambda'}=\omega_B\). Then
	\[
	\sigma_{\lambda'}=q_{\lambda'}\widetilde\rho_{\lambda'}
	\quad
	\text{for every }\lambda'\in\Lambda_D.
	\]
	Hence \(\widetilde\rho_{\lambda'}\) is a state of Bob's system for every
	\(\lambda'\in\Lambda_D\).
		Substituting this identity into the decomposition \eqref{SDD} of
	\(\rho^B_{a|x}\), we obtain
$$\rho^B_{a|x}= \sum_{\lambda'\in\Lambda_D}
	q_{\lambda'}D(a|x,\lambda')\widetilde\rho_{\lambda'}, \ \forall x\in[m_A], a\in[o_A].$$
	This is the finite-sum LHS model in
	Eq.~\eqref{UD-LHSM}. 	
\end{proof}

\noindent\textbf{Lemma 2.}
Let \(\rho_{AB}\) be a state on
\(\mathcal H_A\otimes\mathcal H_B\), and \(T\) be an infinity index set. For each \(t\in T\), Alice
performs a two-outcome POVM
$\mathsf M_t=\{M_{1|t}=M_t,M_{0|t}=I_A-M_t\}$ with $0\le M_t\le I_A$ and induce an assemblage $\Sigma(T)=(\sigma_{a|t})_{t\in T,a=0,1}$ of Bob's side, where
$$\sigma_{a|t}={\rm{Tr}}_A[(M_{1|t}\otimes I_B)\rho_{AB}](t\in T,a=0,1).$$
Assume that the assemblage $\Sigma(T)$ does not admit a common LHS model. Then there exists a finite subset
\(F=\{t_1,\ldots,t_m\}\subseteq T\)  such that the assemblage $\Sigma(F)=(\sigma_{a|t})_{t\in F,a=0,1}$
 does not admit a common LHS model and therefor the state \(\rho_{AB}\) is steerable from Alice to Bob under finitely many two-outcome POVM measurements $\mathsf M_{t_1},\mathsf M_{t_2},\ldots,\mathsf M_{t_m}$.

\begin{proof} Suppose by contradiction that for any finite subset
\(F=\{t_1,\ldots,t_m\}\subseteq T\),  the assemblage $\Sigma(F)=(\sigma_{a|t})_{t\in F,a=0,1}$
  admits a common LHS model.

The proof will be completed according the following $8$ steps.

(1) Define $X=\Omega\times\mathcal{S}_B$, where
  	\begin{equation}
		\Omega=\{0,1\}^{T}=\{\alpha| \alpha: T\longrightarrow\{0,1\}\}=\prod_{t\in T}\{0,1\},
		\label{Omega-def}
	\end{equation}
which is the product space of compact Hausdorff spaces $Y_t=\{0,1\}$ with the discrete topology and then a compact Hausdorff space, and $\mathcal S_B$ is the set of all states of system $B$,
which is a compact Hausdorff space with the usual operator-norm topology. Using Tychonoff's theorem,  $X=\Omega\times\mathcal{S}_B$ becomes a compact Hausdorff space.

(2) Let $B(X)$ 	be the Borel \(\sigma\)-algebra on \(X\), i.e., the smallest \(\sigma\)-algebra containing all open sets in the topology. 	Denote the set of all Radon probability measures on  $B(X)$  by \(\mathcal P(X)\). With the weak*-topology on \(\mathcal P(X)\), i.e.,  the usual narrow topology defined by
$$\mu_\alpha\rightharpoonup\mu\Leftrightarrow\int_X f(x) d\mu_\alpha(x)\to \int_X f(x) d\mu(x), \forall f\in C(X),$$ \(\mathcal P(X)\) becomes a topological space.

(3)	Let \(C(X)\) denote the unital commutative $C^*$-algebra consisting of all real-valued continuous functions on \(X\), equipped with the uniform norm 	\(\|f\|_\infty=\sup_{x\in X}|f(x)|\), and let
	 \(C(X)^{*}\) be the conjugate space consisting of all bounded linear functionals on  \(C(X)\). By the Riesz--Markov--Kakutani representation theorem, every \(L\in C(X)^{*}\) corresponds uniquely to a finite Radon signed measure \(\mu\) such that
	\begin{equation}
		L(f)=\int_X f\,d\mu, \qquad \forall f\in C(X).
		\label{L-mu}
	\end{equation}
Let $\mathcal K$ be the state space   of $C(X)$, i.e., the of all bounded positive linear functionals $f$ on $C(X)$ such that $L(1)=1$. Clearly, $\mathcal K\subset C(X)^*$. By Banach-Alaoglu theorem  in functional analysis, the state space $K$ of  \(C^*\)-algebra is compact in the weak*-topology $\sigma(C(X)^*,C(X))$.
Define a map $\Gamma:\mu\mapsto L_\mu$ from 		
$\mathcal P(X)$ into $\mathcal K$ by
	\begin{equation}
L_\mu(f)=\int_Xf\,d\mu,\ \ \forall \mu\in\mathcal P(X).
		\label{PX}
	\end{equation}
	By the Riesz--Markov--Kakutani representation theorem, \(\mathcal P(X)\)
is homeomorphic to \(\mathcal{K}\) under $\Gamma$. Hence, \(\mathcal P(X)\) is compact in its weak*- topology.

(4)  For each \(t\in T\), define the matrix-valued functions
	\(f_t(\alpha,\rho)=\alpha(t)\rho\),
	and
	\(g(\alpha,\rho)=\rho\).
	Since the coordinate map \(\alpha\mapsto\alpha(t)\) is continuous in the product topology and \((\alpha,\rho)\mapsto\rho\) is a continuous coordinate projection, every matrix element of \(f_t\) and \(g\) is a continuous function on \(X\). Let $\mathcal{C}_t$ be the set of all measures $\mu\in\mathcal P(X)$ such that
	\begin{equation}
		\int_X\alpha(t)\rho\,d\mu(\alpha,\rho)
		=\sigma_{1|t} \mathrm{\,\  and\,  }
		\int_X\rho\,d\mu(\alpha,\rho)=\rho_B,
		\label{Ct-def}
	\end{equation}
where matrix integrals are defined entry-wise.
 If $\mu\in\mathcal{C}_t$, then for each $\lm=(\alpha,\rho)\in X$,
	set
	$$P(1|t,\lm)=\alpha(t),P(0|t,\lm)=1-\alpha(t),\tau_\lm=\rho,$$ hence,
	the LHS model
$$\sigma_{a|t}=\int_XP(a|t,\lm)\tau_\lm\,d\mu(\alpha,\rho)(a=0,1)$$
 given by $\mu$ reproduce the assemblage $\{\sigma_{1|t},\sigma_{0|t}\}$.
	
(5) Let \(\operatorname{Herm}(\mathcal H_B)\) be the finite-dimensional real vector space of all Hermitian matrices on Bob's subsystem, equipped with the usual matrix topology. For each fixed \(t\in T\), define a  map $\Phi_t$ from $\mathcal P(X)$ into $\operatorname{Herm}(\mathcal H_B)\times\operatorname{Herm}(\mathcal H_B)$ by
	\begin{equation}
		\mu\longmapsto \left(
			\int_X\alpha(t)\rho\,d\mu,
			\int_X\rho\,d\mu
			\right).
			\label{Phi_t-def}
	\end{equation}
It can be proved that the map \(\Phi_t\) is continuous.
	
(6)	For each $t\in T$, define the singleton set in the target space \(\operatorname{Herm}(\mathcal H_B)\):
	\begin{equation}
		A_t=\{(\sigma_{1|t},\rho_B)\}.
		\label{At-def}
	\end{equation}
	By the definition of \(\mathcal C_t\),
	\begin{equation}
		\mathcal C_t=\Phi_t^{-1}
		\big(\{(\sigma_{1|t},\rho_B)\}\big).
		\label{Ct-pre}
	\end{equation}
	Since the target space
	\(\operatorname{Herm}(\mathcal H_B)\times\operatorname{Herm}(\mathcal H_B)\)
	is a finite-dimensional Hausdorff space, the singleton set \(A_t\) is closed. Since \(\Phi_t\) is continuous, the preimage of a closed set under a continuous map is closed. Therefore,
	\begin{equation}
	\forall t\in T,\, \,	\mathcal C_t \text{ is a weak*-closed set in } \mathcal P(X).
		\label{Ct-closed}
	\end{equation}
	
(7) For any finitely many distinct measurement parameters
	\(F=\{t_1,\ldots,t_m\}\subseteq T\), define
$\Sigma(F)=(\sigma_{a|t})_{t\in F,a=0,1}$ of Bob's side, where
$$\sigma_{a|t}={\rm{Tr}}_A[(M_{a|t}\otimes I_B)\rho_{AB}](t\in F,a=0,1).$$
	we prove that
	\begin{equation}
		\bigcap_{i=1}^{m}\mathcal C_{t_i} \neq\varnothing
		\iff  \Sigma(F)\text{\, admits an LHS model}.
		\label{L-1}
	\end{equation}
	
	\textbf{Necessity}: Assume there exists
	\(\mu\in\bigcap_{i=1}^{m}\mathcal C_{t_i}\).
	Define the response functions
	\(p(1|t_i,\alpha,\rho)=\alpha(t_i)\) and \(p(0|t_i,\alpha,\rho)=1-\alpha(t_i)\).
	Since \(\alpha(t_i)\in\{0,1\}\), these are valid probability response functions. By Eq.~\eqref{Ct-def}, for each $i=1,2,\ldots,m$,
$$
	\int_Xp(1|t_i,\alpha,\rho)\rho\,d\mu=\sigma_{1|t_i},$$$$
	\int_Xp(0|t_i,\alpha,\rho)\rho\,d\mu
	=\rho_B-\sigma_{1|t_i}=\sigma_{0|t_i}.
	$$
This gives an LHS model for  $\Sigma(F)$.
	
	\textbf{Sufficiency}: Assume that  $\Sigma(F)$ admits an LHS model. Thus, for some probability space
	\((\Lambda_F,\Sigma_F,\nu_F)\), normalized hidden states
	\(\rho_\lambda\), and response probabilities \(P(a|i,\lambda)\), it holds that
	\begin{equation}
		\sigma_{a|t_i}
		=
		\int_{\Lambda_F} P(a|i,\lambda)\rho_\lambda\,d\nu_F(\lambda)
			\label{LHS-model}
	\end{equation}
for all $a\in\{0,1\}, i\in[m].$ According to Lemma 1, \eqref{LHS-model} can be written as
\begin{equation}\label{C-2}
		\sigma_{a|t_i}=\sum_{\lm'\in \Lambda_D}q_{\lambda'}D(a|t_i,\lm')\widetilde\rho_{\lambda'},\  \forall i\in[m],a=0,1,
\end{equation}
where $\Lambda_D$ denotes the set of all maps from $[m]$ onto $\{0,1\}$, $\rho_{\lambda'}$'s are unnormalized states of Bob's system, $q_{\lambda'}$ and $D(a|t_i,\lm')=\delta_{a,\lm'(x)}$ are probability distributions of $\lm'$ and $a$.
	Extend each finite deterministic strategy $\lm'$  to an element $\alpha_{\lambda'}$ of
	\(\Omega=\{0,1\}^T\) by setting
	\begin{equation}
		\alpha_{\lambda'}(t)
		=
		\begin{cases}
			\lambda'(i), & t=t_i,\quad i\in[m],\\
			0, & t\notin F,
		\end{cases}
		\label{D-1}
	\end{equation}
and then define a probability measure
	\begin{equation}
		\mu_F=\sum_{\lambda'\in\Lambda_D}
		q_{\lambda'}\,
		\delta_{(\alpha_{\lambda'},\widetilde\rho_{\lambda'})}.
		\label{measure}
	\end{equation}
	This is a finite convex combination of Dirac measures $\delta_{(\alpha_{\lambda'},\widetilde\rho_{\lambda'})}$ on \(X\);
	hence \(\mu_F\in\mathcal P(X)\). 	
	By construction of $\mu_F$, we get that
	\begin{equation}
		\int_X \rho\,d\mu_F(\alpha,\rho)
		=
		\sum_{\lambda'}q_{\lambda'}\widetilde\rho_{\lambda'}
		=
		\sum_{\lambda'}\sigma_{\lambda'}
		=
		\rho_B,
		\label{D-2}
	\end{equation}
and for every \(i=1,\ldots,m\), we compute from Eqs.~\eqref{measure} and \eqref{C-2} that
	\begin{eqnarray*}
\int_X\alpha(t_i)\rho\,d\mu_F(\alpha,\rho)
&=&\sum_{\lambda'}q_{\lambda'}\alpha_{\lambda'}(t_i)\widetilde\rho_{\lambda'}\\
&=&\sum_{\lambda'}D(1|t_i,\lambda')\sigma_{\lambda'}\\
& =&\sigma_{1|t_i}.
	\end{eqnarray*}
Thus \(\mu_F\in\mathcal C_{t_i}\) for every \(i=1,\ldots,m\), and
	therefore
	\[
	\bigcap_{i=1}^m\mathcal C_{t_i}\neq\varnothing.
	\]
	
(8) Since it was assumed that for any finite subset
	\(F=\{t_1,\ldots,t_m\}\subset T\),
	the corresponding assemblage $\Sigma(F)$ admits a common LHS model, we see from  Eq.~\eqref{L-1} that
	\(\bigcap_{i=1}^{m}\mathcal C_{t_i} \neq\varnothing\).
 the family of closed sets
	\(\{\mathcal C_t:t\in T\}\) in the compact space \(\mathcal P(X)\)
	has the finite intersection property. Hence, $\bigcap_{t\in T}\mathcal C_t \neq\varnothing$ 	by the finite intersection property of compact spaces.
	Therefore, there exists a common Radon probability measure $\mu\in \bigcap_{t\in T}\mathcal C_t \neq\varnothing$. By the definition of $\mathcal C_t$, we know that for all \(t\in T\),
	\begin{equation}
		\int_X\alpha(t)\rho\,d\mu(\alpha,\rho) =\sigma_{1|t},\quad
		\int_X\rho\,d\mu(\alpha,\rho)=\rho_B.
	\end{equation}
	Define the response functions
	\(p(1|t,\alpha,\rho)=\alpha(t)\), \(p(0|t,\alpha,\rho)=1-\alpha(t)\).
	Since the coordinate map \(\alpha\mapsto\alpha(t)\) is continuous in $(\alpha,\rho)$, the response functions are measurable with respect to the hidden variable  $(\alpha,\rho)$. Thus
	\begin{equation}
		\sigma_{a|t} = \int_X p(a|t,\alpha,\rho)\rho\, d\mu(\alpha,\rho),
		\quad a=0,1,\quad t\in T.
	\end{equation}
	This constructs a common LHS model for the assemblage $\Sigma(T)$, contradicting the initial assumption of the lemma. This completes the proof.
\end{proof}

\noindent\textbf{Theorem 2 (Geometry criterion).}
\label{thm:asymptotic_steering_SM} {\it Suppose that there exist constants \(L>0\), \(C>0\), and \(t_1\in(0,t_0)\) such that, for all \(0<t<t_1\),
	\begin{equation}
		T_t\geq Lt,
		\qquad
		D_t\leq Ct^2.
	\end{equation}
	Then the family of conditional states \(\{\sigma_t:0<t<t_1\}\) does not admit a common LHS model, i.e., the family  \(\{\sigma_t:0<t<t_1\}\) does not admit a common LHS model, i.e., there does not exist a probability measure space \((\Lambda,\Sigma,\mu)\),  a measurable family of density operators \(\{\tau_\lambda\}_{\lambda\in\Lambda}\),  and measurable response functions \(P_t:\Lambda\to[0,1]\) such that
	\begin{equation}\label{s-THM2}
		\sigma_t
		=
		\int_{\Lambda}
		P_t(\lambda)\tau_\lambda\,d\mu(\lambda),
		\qquad
		\forall 0<t<t_1.
		%\label{eq:lhs-model}
	\end{equation}
Consequently, the corresponding two-qutrit state \(\rho_{AB}\) is steerable from Alice to Bob under finitely many of two-outcome POVM measurements  $\mathsf M_{t_1},\mathsf M_{t_2},\ldots,\mathsf M_{t_m}$.}

\begin{proof}
	Assume, for contradiction, that the family admits a common LHS model \eqref{s-THM2}. 	
	For every \(\lambda\in\Lambda\), write \(\tau_\lambda=\frac{1}{3}(I_3+\vec r(\lambda)\cdot\vec\delta)\), where \(\vec r(\lambda)=(r_1(\lambda),\ldots,r_8(\lambda))\) is the Bloch vector of \(\tau_\lambda\). Define \(p_j(\lambda):=\bra{j}\tau_\lambda\ket{j}\) for \(j=0,1,2\), and
	\begin{equation}
		q(\lambda)
		:=
		p_1(\lambda)+p_2(\lambda)
		=
		\bra{1}\tau_\lambda\ket{1}
		+
		\bra{2}\tau_\lambda\ket{2}.
		\label{eq:q-definition}
	\end{equation}
	Since \(\tau_\lambda\) is a density operator, \(p_0(\lambda)+p_1(\lambda)+p_2(\lambda)=1\).
	
	Also define the four-dimensional transverse vector
	\begin{equation}
		u(\lambda)
		:=
		\bigl(
		r_1(\lambda),r_2(\lambda),
		r_4(\lambda),r_5(\lambda)
		\bigr).
		\label{eq:u-definition}
	\end{equation}
	Using the Bloch-coordinate relations,
	\[
	u(\lambda)=
	\sqrt{3}\bigl(
	\operatorname{Re}\bra{0}\tau_\lambda\ket{1},
	-\operatorname{Im}\bra{0}\tau_\lambda\ket{1},
	\operatorname{Re}\bra{0}\tau_\lambda\ket{2},
	-\operatorname{Im}\bra{0}\tau_\lambda\ket{2}
	\bigr).
	\]
	Because \(\tau_\lambda\ge0\), every \(2\times2\) principal minor is nonnegative. Hence
	\begin{equation}
		\left|
		\bra{i}\tau_\lambda\ket{j}
		\right|^2
		\leq
		\bra{i}\tau_\lambda\ket{i}
		\bra{j}\tau_\lambda\ket{j},
		\qquad
		i,j\in\{0,1,2\}.
		\label{eq:psd-minor-inequality}
	\end{equation}
	It follows that
	\begin{align}
		\|u(\lambda)\|_2^2
		=
		3\left(
		\left|\bra{0}\tau_\lambda\ket{1}\right|^2
		+
		\left|\bra{0}\tau_\lambda\ket{2}\right|^2
		\right)
		\leq
		3p_0(\lambda)
		\bigl(p_1(\lambda)+p_2(\lambda)\bigr)
		=
		3p_0(\lambda)q(\lambda)
		\leq
		3q(\lambda).
	\end{align}
	Therefore,
	\begin{equation}
		\|u(\lambda)\|_2
		\leq
		\sqrt{3q(\lambda)}.
		\label{eq:pointwise-transverse-bound}
	\end{equation}
	
	Next, taking the \((1,1)\) and \((2,2)\) matrix elements of Eq.~\eqref{eq:lhs-model}, we obtain
	\begin{align}
		D_t
		=
		d_t+f_t
		=
		\int_{\Lambda}
		P_t(\lambda)
		\left(
		\bra{1}\tau_\lambda\ket{1}
		+
		\bra{2}\tau_\lambda\ket{2}
		\right)
		d\mu(\lambda)
		=
		\int_{\Lambda}
		P_t(\lambda)q(\lambda)\,d\mu(\lambda)
		\leq Ct^2.
		\label{eq:integrated-defect-bound}
	\end{align}
	Taking instead the \((0,1)\) and \((0,2)\) matrix elements of Eq.~\eqref{eq:lhs-model}, and using the definition of \(u(\lambda)\), gives
	\[
	\int_{\Lambda}
	P_t(\lambda)u(\lambda)\,d\mu(\lambda)
	=
	\sqrt{3}\bigl(
	\operatorname{Re}(b_t),-\operatorname{Im}(b_t),
	\operatorname{Re}(c_t),-\operatorname{Im}(c_t)
	\bigr).
	\]
	Consequently,
	\begin{equation}
		\left\|
		\int_{\Lambda}
		P_t(\lambda)u(\lambda)\,d\mu(\lambda)
		\right\|_2
		=
		\sqrt{3\bigl(|b_t|^2+|c_t|^2\bigr)}
		=
		T_t.
		\label{eq:integrated-transverse-vector}
	\end{equation}
	By the assumed lower bound on \(T_t\),
	\begin{equation}
		\left\|
		\int_{\Lambda}
		P_t(\lambda)u(\lambda)\,d\mu(\lambda)
		\right\|_2
		\geq Lt.
		\label{eq:total-transverse-lower-bound}
	\end{equation}
	
	Choose a constant \(K\) satisfying
	\begin{equation}
		K\geq\frac{2\sqrt{3}\,C}{L},
		\label{eq:choice-of-K}
	\end{equation}
	and define
	\begin{equation}
		\Gamma_t
		:=
		\left\{
		\lambda\in\Lambda:
		q(\lambda)\leq K^2t^2
		\right\}.
		\label{eq:gamma-definition}
	\end{equation}
	On the complement \(\Gamma_t^c\), one has \(q(\lambda)>K^2t^2\), hence \(\sqrt{q(\lambda)}<q(\lambda)/(Kt)\). Using Eq.~\eqref{eq:pointwise-transverse-bound} and Eq.~\eqref{eq:integrated-defect-bound}, we obtain
	\begin{align}
		\left\|
		\int_{\Gamma_t^c}
		P_t(\lambda)u(\lambda)\,d\mu(\lambda)
		\right\|_2
		&\leq
		\int_{\Gamma_t^c}
		P_t(\lambda)\|u(\lambda)\|_2\,d\mu(\lambda)
		\leq
		\frac{\sqrt{3}}{Kt}
		\int_{\Gamma_t^c}
		P_t(\lambda)q(\lambda)\,d\mu(\lambda)
		\notag\\
		&\leq
		\frac{\sqrt{3}}{Kt}
		\int_{\Lambda}
		P_t(\lambda)q(\lambda)\,d\mu(\lambda)\leq
		\frac{\sqrt{3}C}{K}t
		\leq
		\frac{L}{2}t.
		\label{eq:outer-transverse-bound}
	\end{align}
	
	Let
	\begin{equation}
		\Lambda_0
		:=
		\left\{
		\lambda\in\Lambda:q(\lambda)=0
		\right\}.
		\label{eq:lambda-zero}
	\end{equation}
	If \(\lambda\in\Lambda_0\), then \(p_1(\lambda)=p_2(\lambda)=0\). By Eq.~\eqref{eq:psd-minor-inequality}, all matrix elements of \(\tau_\lambda\) involving \(\ket{1}\) or \(\ket{2}\) must vanish. Since \(\operatorname{Tr}(\tau_\lambda)=1\), it follows that
	\begin{equation}
		\tau_\lambda=\ket{0}\!\bra{0},\qquad
		\forall\lambda\in\Lambda_0.
		\label{eq:zero-defect-state}
	\end{equation}
	In particular,
	\begin{equation}
		u(\lambda)=0,
		\qquad
		\forall\lambda\in\Lambda_0.
		\label{eq:zero-transverse-vector}
	\end{equation}
	
	On \(\Gamma_t\), we have \(q(\lambda)\leq K^2t^2\). Thus, by Eq.~\eqref{eq:pointwise-transverse-bound}, \(\|u(\lambda)\|_2\leq\sqrt{3}\,Kt\) for \(\lambda\in\Gamma_t\). Using \(0\leq P_t(\lambda)\leq1\) and Eq.~\eqref{eq:zero-transverse-vector}, we find
	\begin{align}
		\left\|
		\int_{\Gamma_t}
		P_t(\lambda)u(\lambda)\,d\mu(\lambda)
		\right\|_2
		\leq
		\int_{\Gamma_t}
		P_t(\lambda)\|u(\lambda)\|_2\,d\mu(\lambda)
		\leq
		\sqrt{3}\,Kt\,
		\mu\bigl(
		\Gamma_t\setminus\Lambda_0
		\bigr).
		\label{eq:inner-transverse-upper-bound}
	\end{align}
	
	On the other hand, decompose the total transverse vector as
	\[
	\int_{\Lambda}
	P_t(\lambda)u(\lambda)\,d\mu(\lambda)
	=
	\int_{\Gamma_t}
	P_t(\lambda)u(\lambda)\,d\mu(\lambda)
	+
	\int_{\Gamma_t^c}
	P_t(\lambda)u(\lambda)\,d\mu(\lambda).
	\]
	The reverse triangle inequality, together with Eqs.~\eqref{eq:total-transverse-lower-bound} and~\eqref{eq:outer-transverse-bound}, gives
	\begin{align}
		\left\|
		\int_{\Gamma_t}
		P_t(\lambda)u(\lambda)\,d\mu(\lambda)
		\right\|_2
		\geq
		\left\|
		\int_{\Lambda}
		P_t(\lambda)u(\lambda)\,d\mu(\lambda)
		\right\|_2
		-
		\left\|
		\int_{\Gamma_t^c}
		P_t(\lambda)u(\lambda)\,d\mu(\lambda)
		\right\|_2
		\geq
		Lt-\frac{L}{2}t
		=
		\frac{L}{2}t.
		\label{eq:inner-transverse-lower-bound}
	\end{align}
	Combining Eqs.~\eqref{eq:inner-transverse-upper-bound} and~\eqref{eq:inner-transverse-lower-bound}, we obtain
	\begin{equation}
		\mu\bigl(
		\Gamma_t\setminus\Lambda_0
		\bigr)
		\geq
		\frac{L}{2\sqrt{3}\,K}
		>0.
		\label{eq:uniform-measure-lower-bound}
	\end{equation}
	
	Finally, let \(t_n\downarrow0\) be any strictly decreasing sequence with \(t_n<t_1\), and define $$A_n:=\Gamma_{t_n}\setminus\Lambda_0=\{\lambda\in\Lambda:0<q(\lambda)\leq K^2t_n^2\}.$$ Then \(A_{n+1}\subseteq A_n\) and \(\bigcap_{n=1}^{\infty}A_n=\varnothing\). Since \(\mu\) is a probability measure, it is finite. Therefore, by continuity from above,
	\begin{equation}
		\lim_{n\to\infty}\mu(A_n)
		=
		\mu\left(
		\bigcap_{n=1}^{\infty}A_n
		\right)
		=
		0.
		\label{eq:measure-continuity}
	\end{equation}
	This contradicts the uniform positive lower bound in Eq.~\eqref{eq:uniform-measure-lower-bound}. Hence, the assumed common LHS model cannot exist. Consequently, the bigger family $\Sigma(T)=(\sigma_{a|t})_{t\in T,a=0,1}$ with $T=(0,t_0)$ does not admit a LHS model. 
According to Lemma 2, we conclude that  the state \(\rho_{AB}\) is steerable from Alice to Bob under finitely many two-outcome POVM measurements  $\mathsf M_{t_1},\mathsf M_{t_2},\ldots,\mathsf M_{t_m}$.
\end{proof}

\subsection{Relation between Theorems~1 and~2}
\label{app:sec:connection_theorems}

This section establishes the relation between Theorems~1 and~2 through two results, proved in the two subsections below.

\subsubsection{From Theorem~1 states to Theorem~2}
\label{app:subsec:theorem1_to_theorem2}

\textbf{Proposition 1}
\label{prop:orthogonalization}
Let \(\rho_{AB}\) be a two-qutrit state satisfying the structural
assumptions of Theorem~1, i.e.,
\begin{align}
	\rho_{AB}
	={}
	\sum_{j=0}^{2}\mu_{j+1}
	|j\rangle\langle j|
	\otimes|\phi_{j+1}\rangle\langle\phi_{j+1}|
	+
	\bigl(
	|0\rangle\langle1|\otimes\mathcal A
	+
	|0\rangle\langle2|\otimes\mathcal B
	+
	|1\rangle\langle2|\otimes\mathcal C
	+\mathrm{H.c.}
	\bigr).
	\label{eq:theory_pure_conditional}
\end{align}
Then there exists an invertible local filter
\(K\) on Bob's subsystem, with \(K^\dagger K\preceq I_B\), such
that
\begin{align}
	\widetilde{\rho}_{AB}
	={}
	\frac{(I_A\otimes K)\rho_{AB}(I_A\otimes K^\dagger)}{p_K}
	={}
	\sum_{j=0}^{2}\tilde{\mu}_{j+1}
	|j\rangle\langle j|
	\otimes|j\rangle\langle j|
	+
	\bigl(
	|0\rangle\langle1|\otimes\tilde{\mathcal A}
	+
	|0\rangle\langle2|\otimes\tilde{\mathcal B}
	+
	|1\rangle\langle2|\otimes\tilde{\mathcal C}
	+\mathrm{H.c.}
	\bigr).
	\label{eq:theory_pure_conditional-1}
\end{align}
%where \(\tilde{\mu}_{j+1}=\mu_{j+1}\) and
%\(\tilde{\mathcal A}=c_{01}|0\rangle\langle1|\),
%\(\tilde{\mathcal B}=c_{02}|0\rangle\langle2|\),
%\(\tilde{\mathcal C}=c_{12}|1\rangle\langle2|\) for some
%\(c_{01},c_{02},c_{12}\in\mathbb{C}\).

\begin{proof}
	Let \(\Phi=(|\phi_1\rangle,|\phi_2\rangle,|\phi_3\rangle)\) be the
	matrix whose columns are the conditional vectors. Linear
	independence makes \(\Phi\) invertible. Choose
	\begin{equation}
		K=\eta\Phi^{-1},
		\qquad
		0<\eta\leq\frac{1}{\|\Phi^{-1}\|_{\mathrm{op}}}.
		\label{eq:link_filter}
	\end{equation}
	Then \(K^\dagger K\preceq I_B\) and
	\(K|\phi_{i+1}\rangle=\eta|i\rangle\) for \(i=0,1,2\).
	
	We compute the filtered state directly from
	Eq.~\eqref{eq:theory_pure_conditional}. Write
	\begin{equation}
		\rho_{AB}
		=
		\sum_{i,j=0}^{2}|i\rangle\langle j|\otimes R_{ij},
		\quad
		R_{ii}=\mu_{i+1}|\phi_{i+1}\rangle\langle\phi_{i+1}|.
		\label{eq:block_form}
	\end{equation}
	By \(\rho_{AB}\ge0\) and the purity of the diagonal blocks,
	positivity forces each off-diagonal block to have the form
	\begin{equation}
		R_{ij}
		=
		c_{ij}|\phi_{i+1}\rangle\langle\phi_{j+1}|,
		\qquad
		c_{ji}=\overline{c_{ij}},
		\qquad
		c_{ii}=\mu_{i+1},
		\label{eq:link_coherence_blocks}
	\end{equation}
	so in particular \(\mathcal A=c_{01}|\phi_1\rangle\langle\phi_2|\),
	\(\mathcal B=c_{02}|\phi_1\rangle\langle\phi_3|\), and
	\(\mathcal C=c_{12}|\phi_2\rangle\langle\phi_3|\).
	
	Applying the filter \(K\) on Bob's side gives
	\begin{equation}
		(I_A\otimes K)\rho_{AB}(I_A\otimes K^\dagger)
		=
		\sum_{i,j=0}^{2}|i\rangle\langle j|\otimes K R_{ij}K^\dagger.
		\label{eq:filtered_blocks}
	\end{equation}
	We evaluate each block separately. For the diagonal blocks, using
	\(K|\phi_{i+1}\rangle=\eta|i\rangle\) and
	\(\langle\phi_{j+1}|K^\dagger=\eta\langle j|\),
	\begin{equation}
		K R_{ii}K^\dagger
		=
		\mu_{i+1}\,
		\bigl(K|\phi_{i+1}\rangle\bigr)
		\bigl(\langle\phi_{i+1}|K^\dagger\bigr)
		=
		\eta^2\mu_{i+1}|i\rangle\langle i|.
		\label{eq:filtered_diagonal_block}
	\end{equation}
	For the off-diagonal blocks,
	\begin{align}
		K\mathcal A K^\dagger
		&=
		c_{01}\,
		\bigl(K|\phi_1\rangle\bigr)
		\bigl(\langle\phi_2|K^\dagger\bigr)
		=
		\eta^2 c_{01}|0\rangle\langle1|,
		\label{eq:filtered_A}\\
		K\mathcal B K^\dagger
		&=
		c_{02}\,
		\bigl(K|\phi_1\rangle\bigr)
		\bigl(\langle\phi_3|K^\dagger\bigr)
		=
		\eta^2 c_{02}|0\rangle\langle2|,
		\label{eq:filtered_B}\\
		K\mathcal C K^\dagger
		&=
		c_{12}\,
		\bigl(K|\phi_2\rangle\bigr)
		\bigl(\langle\phi_3|K^\dagger\bigr)
		=
		\eta^2 c_{12}|1\rangle\langle2|.
		\label{eq:filtered_C}
	\end{align}
	Substituting
	Eqs.~\eqref{eq:filtered_diagonal_block}--\eqref{eq:filtered_C}
	into Eq.~\eqref{eq:filtered_blocks} yields
	\begin{align}
		&(I_A\otimes K)\rho_{AB}(I_A\otimes K^\dagger)
		=
		\eta^2
		\sum_{j=0}^{2}\mu_{j+1}|j\rangle\langle j|\otimes|j\rangle\langle j|
		+
		\eta^2
		\bigl(
		c_{01}|0\rangle\langle1|\otimes|0\rangle\langle1|
		+
		c_{02}|0\rangle\langle2|\otimes|0\rangle\langle2|
		+
		c_{12}|1\rangle\langle2|\otimes|1\rangle\langle2|
		+\mathrm{H.c.}
		\bigr).
		\label{eq:filtered_unnormalized}
	\end{align}
	The probability is
	\begin{equation}
		p_K
		=
		\operatorname{Tr}\!\left[
		(I_A\otimes K)\rho_{AB}(I_A\otimes K^\dagger)
		\right]
		=
		\eta^2\sum_{j=0}^{2}\mu_{j+1}
		=
		\eta^2.
		\label{eq:link_filter_success_probability}
	\end{equation}
	Conditioning on the successful outcome gives
	\begin{align}
		\widetilde{\rho}_{AB}
		={}&
		\frac{(I_A\otimes K)\rho_{AB}(I_A\otimes K^\dagger)}{p_K}
		\nonumber\\
		={}&
		\sum_{j=0}^{2}\mu_{j+1}|j\rangle\langle j|\otimes|j\rangle\langle j|
		+
		\bigl(
		c_{01}|0\rangle\langle1|\otimes|0\rangle\langle1|
		+
		c_{02}|0\rangle\langle2|\otimes|0\rangle\langle2|
		+
		c_{12}|1\rangle\langle2|\otimes|1\rangle\langle2|
		+\mathrm{H.c.}
		\bigr).
		\label{eq:filtered_normalized}
	\end{align}
	Setting
	\begin{equation}
		\tilde{\mu}_{j+1}:=\mu_{j+1},
		\qquad j=0,1,2,
	\end{equation}
	and
	\[
	\tilde{\mathcal A}=c_{01}|0\rangle\langle1|,
	\qquad
	\tilde{\mathcal B}=c_{02}|0\rangle\langle2|,
	\qquad
	\tilde{\mathcal C}=c_{12}|1\rangle\langle2|,
	\]
	we can rewrite Eq.~\eqref{eq:filtered_normalized} as
	\begin{align}
		\widetilde{\rho}_{AB}
		={}
		\sum_{j=0}^{2}\tilde{\mu}_{j+1}
		|j\rangle\langle j|
		\otimes|j\rangle\langle j|
		+
		\bigl(
		|0\rangle\langle1|\otimes\tilde{\mathcal A}
		+
		|0\rangle\langle2|\otimes\tilde{\mathcal B}
		+
		|1\rangle\langle2|\otimes\tilde{\mathcal C}
		+\mathrm{H.c.}
		\bigr),
		\label{eq:theory_pure_conditional-1-proof}
	\end{align}
	which is the stated orthogonal form.

\end{proof}
	
	In the ordered product basis
	\[
	\{|00\rangle,|01\rangle,|02\rangle,
	|10\rangle,|11\rangle,|12\rangle,
	|20\rangle,|21\rangle,|22\rangle\},
	\]
	the state in Eq.~\eqref{eq:theory_pure_conditional-1-proof} takes
	the explicit form
	\begin{equation}
		\widetilde{\rho}_{AB}
		=
		\begin{pmatrix}
			\tilde{\mu}_1 & & & & a_{12} & & & & b_{13}\\
			& 0 & & & & & & & \\
			& & 0 & & & & & & \\
			& & & 0 & & & & & \\
			\overline{a_{12}} & & & & \tilde{\mu}_2 & & & & c_{23}\\
			& & & & & 0 & & & \\
			& & & & & & 0 & & \\
			& & & & & & & 0 & \\
			\overline{b_{13}} & & & & \overline{c_{23}} & & & & \tilde{\mu}_3
		\end{pmatrix},
		\label{eq:link_explicit_matrix}
	\end{equation}
	where blank entries are zero, and
	\(a_{12}=c_{01}\), \(b_{13}=c_{02}\), \(c_{23}=c_{12}\). Its
	nonzero part is the positive semidefinite coefficient matrix
	\begin{equation}
		\Gamma
		=
		\begin{pmatrix}
			\tilde{\mu}_1 & a_{12} & b_{13}\\
			\overline{a_{12}} & \tilde{\mu}_2 & c_{23}\\
			\overline{b_{13}} & \overline{c_{23}} & \tilde{\mu}_3
		\end{pmatrix},
		\label{eq:link_explicit_gamma}
	\end{equation}
	so
	\begin{equation}
		\operatorname{rank}(\widetilde{\rho}_{AB})
		=
		\operatorname{rank}(\Gamma)
		\leq 3.
		\label{eq:link_rank_bound}
	\end{equation}

\textbf{Proposition 2}
\label{prop:theorem2}
Let \(\widetilde{\rho}_{AB}\) be the orthogonal form of Eq.~\eqref{eq:link_explicit_matrix}. If at least one of \(a_{12},b_{13},c_{23}\) is nonzero, then \(\widetilde{\rho}_{AB}\)
satisfies the asymptotic conditions of Theorem~2 and is therefore steerable: there exist constants \(L>0\), \(C>0\), and \(t_1>0\)
such that, for all \(0<t<t_1\),
\[
T_t\geq Lt,
\qquad
D_t\leq Ct^2.
\]

\begin{proof}
	In the computational basis, Eq.~\eqref{eq:link_explicit_matrix}
	is equivalently
	\begin{align}
		\widetilde{\rho}_{AB}
		={}
		\tilde{\mu}_1|00\rangle\langle00|
		+\tilde{\mu}_2|11\rangle\langle11|
		+\tilde{\mu}_3|22\rangle\langle22|
		+
		\bigl(
		a_{12}|00\rangle\langle11|
		+b_{13}|00\rangle\langle22|
		+c_{23}|11\rangle\langle22|
		+\mathrm{H.c.}
		\bigr).
		\label{eq:prop2_ket_form}
	\end{align}
	
	A simultaneous permutation of Alice's and Bob's computational bases
	maps \(|ii\rangle\langle jj|\) to
	\(|\pi(i)\pi(i)\rangle\langle\pi(j)\pi(j)|\) and therefore
	preserves the diagonal-plus-coherence structure of
	Eq.~\eqref{eq:prop2_ket_form}. We choose the permutation according
	to which coefficient is nonzero: Take the identity permutation if \(a_{12}\neq0\); take
	\(\pi(0)=0,\ \pi(1)=2,\ \pi(2)=1\) if \(a_{12}=0\) and
	\(b_{13}\neq0\); and take \(\pi(0)=2,\ \pi(1)=0,\ \pi(2)=1\) if
	\(a_{12}=b_{13}=0\), in which case \(c_{23}\neq0\).
	Define
	\[
	U_\pi=\sum_{i=0}^{2}|\pi(i)\rangle\langle i|,
	\qquad
	\widehat{\rho}_{AB}
	=
	(U_\pi\otimes U_\pi)\widetilde{\rho}_{AB}
	(U_\pi^\dagger\otimes U_\pi^\dagger).
	\]
	In the three cases, the new \(01\) coefficient is respectively
	\(a_{12}\), \(b_{13}\), and \(c_{23}\), so it is nonzero. We write
	\begin{equation}
		\widehat{\rho}_{AB}
		=
		\sum_{i,j=0}^{2}\widehat c_{ij}|ii\rangle\langle jj|,
		\qquad
		\widehat c_{01}\neq0,
		\label{eq:link_nonzero_01}
	\end{equation}
	and set \(\nu_i=\widehat c_{ii}\) for \(i=0,1,2\). Positivity of
	\(\widehat{\rho}_{AB}\) implies positivity of its compression to
	\(\operatorname{span}\{|00\rangle,|11\rangle\}\):
	\[
	\begin{pmatrix}
		\nu_0 & \widehat c_{01}\\
		\overline{\widehat c_{01}} & \nu_1
	\end{pmatrix}
	\ge0.
	\]
	Consequently,
	\[
	|\widehat c_{01}|^2\leq\nu_0\nu_1,
	\]
	so \(\nu_0,\nu_1>0\).
	
	Let \(t_0>0\) be the parameter bound in Theorem~2. For
	\(0<t<t_0\), consider Alice's rank-one projective measurement
	effect
	\begin{equation}
		M_t=|v_t\rangle\langle v_t|,
		\qquad
		|v_t\rangle=\frac{|0\rangle+t|1\rangle}{\sqrt{1+t^2}}.
		\label{eq:link_measurement_family}
	\end{equation}
	Since \(\langle v_t|2\rangle=0\), only the \(0\) and \(1\) levels
	contribute to the conditional state. The corresponding
	unnormalized conditional state on Bob is
	\begin{align}
		\sigma_t
		=
		\operatorname{Tr}_A[
		(M_t\otimes I_B)\widehat{\rho}_{AB}]
		=
		\frac{1}{1+t^2}
		\begin{pmatrix}
			\nu_0 & t\widehat c_{01} & 0\\
			t\overline{\widehat c_{01}} & t^2\nu_1 & 0\\
			0&0&0
		\end{pmatrix}.
		\label{eq:link_conditional_state}
	\end{align}
	Writing
	\[
	\sigma_t=
	\begin{pmatrix}
		a_t&b_t&c_t\\
		\overline{b_t}&d_t&e_t\\
		\overline{c_t}&\overline{e_t}&f_t
	\end{pmatrix},
	\]
	we have
	\[
	b_t=\frac{t\widehat c_{01}}{1+t^2},\qquad
	c_t=0,\qquad
	d_t=\frac{t^2\nu_1}{1+t^2},\qquad
	f_t=0.
	\]
	Thus the quantities in Theorem~2 obey
	\begin{equation}
		T_t=\sqrt{3(|b_t|^2+|c_t|^2)}
		=\frac{\sqrt{3}\,|\widehat c_{01}|\,t}{1+t^2},
		\qquad
		D_t=d_t+f_t
		=\frac{\nu_1t^2}{1+t^2}.
		\label{eq:link_TD}
	\end{equation}
	Choose \(t_1\) such that \(0<t_1<\min\{t_0,1\}\). Then for
	\(0<t<t_1\), \(1/(1+t^2)\geq1/2\), and hence
	\begin{equation}
		T_t\geq Lt,
		\qquad
		D_t\leq Ct^2,
		\qquad
		L=\frac{\sqrt{3}}{2}|\widehat c_{01}|>0,
		\qquad
		C=\nu_1>0.
		\label{eq:link_theorem2_bounds}
	\end{equation}
	The positivity of \(L\) follows from \(\widehat c_{01}\neq0\), and
	the positivity of \(C\) follows from \(\nu_1>0\). These are the
	asymptotic conditions of Theorem~2, so \(\widehat{\rho}_{AB}\) is
	steerable. Since \(\widehat{\rho}_{AB}\) and \(\widetilde{\rho}_{AB}\)
	differ by a simultaneous local unitary, which preserves
	steerability, \(\widetilde{\rho}_{AB}\) is steerable as well.
\end{proof}

Propositions~1 and~2
together show that every Theorem~1 steerable state can be mapped, by an invertible local filter, to a state that satisfies the asymptotic conditions of
Theorem~2. In this sense, Theorem~2 covers the entire steerable class of Theorem~1, while Theorem~2 also applies to states outside
that class (as shown in~\ref{app:subsec:theorem2_outside_theorem1}).

\subsubsection{Theorem~2 states outside the Theorem~1 class}
\label{app:subsec:theorem2_outside_theorem1}

In this subsection, we show that Theorem~2 extends beyond the
Theorem~1 class by giving a rank-five state family that satisfies
Theorem~2 but not Theorem~1.

Let
\begin{equation}
	\rho_{\mathrm{T2}}
	=
	p_1\ket{\psi(\theta,\phi)}\bra{\psi(\theta,\phi)}
	+p_2\ket{10}\bra{10}
	+p_3\ket{12}\bra{12}
	+p_4\ket{20}\bra{20}
	+p_5\ket{21}\bra{21},
	\label{eq:link_T2_state}
\end{equation}
where
\begin{equation}
	\ket{\psi(\theta,\phi)}
	=
	\sin\theta\cos\phi\,\ket{00}
	+
	\sin\theta\sin\phi\,\ket{11}
	+
	\cos\theta\,\ket{22},
	\label{eq:link_T2_pure_state}
\end{equation}
and \(p_i>0\), \(\sum_{i=1}^{5}p_i=1\), \(0<\theta,\phi<\pi/2\).
The five pure states in Eq.~\eqref{eq:link_T2_state} are mutually
orthogonal, and \(\ket{\psi(\theta,\phi)}\) is normalized. Thus,
\(\rho_{\mathrm{T2}}\) is a normalized quantum state of rank five.

\paragraph{The state satisfies Theorem~2.}
For Alice's projective measurement defined by
\begin{equation}
	P_t=\ket{\xi_t}\bra{\xi_t},
	\qquad
	\ket{\xi_t}
	=
	\frac{\ket{0}+t\ket{1}}{\sqrt{1+t^2}},
	\qquad t>0,
	\label{eq:link_T2_measurement}
\end{equation}
the unnormalized conditional state on Bob's subsystem is
\begin{equation}
	\sigma_t
	=
	\frac{1}{1+t^2}
	\begin{pmatrix}
		p_1\sin^2\theta\cos^2\phi+t^2p_2
		&
		tp_1\sin^2\theta\cos\phi\sin\phi
		&
		0
		\\
		tp_1\sin^2\theta\cos\phi\sin\phi
		&
		t^2p_1\sin^2\theta\sin^2\phi
		&
		0
		\\
		0&0&t^2p_3
	\end{pmatrix}.
	\label{eq:link_T2_conditional_state}
\end{equation}
Consequently,
\begin{equation}
	T_t
	=
	\frac{
		\sqrt{3}\,p_1\sin^2\theta\cos\phi\sin\phi\,t
	}{
		1+t^2
	},
	\qquad
	D_t
	=
	\frac{
		\left(
		p_1\sin^2\theta\sin^2\phi+p_3
		\right)t^2
	}{
		1+t^2
	}.
	\label{eq:link_T2_transverse_leakage}
\end{equation}
Since \(p_1>0\) and \(0<\theta,\phi<\pi/2\), the coefficient
\(p_1\sin^2\theta\cos\phi\sin\phi\) is strictly positive. For
\(0<t<\min\{t_0,1\}\), where \(t_0\) is the bound in Theorem~2,
\[
T_t
\geq
\frac{\sqrt{3}}{2}
p_1\sin^2\theta\cos\phi\sin\phi\,t,
\qquad
D_t
\leq
\left(
p_1\sin^2\theta\sin^2\phi+p_3
\right)t^2.
\]
The constants on the right-hand sides are strictly positive, so
\(\rho_{\mathrm{T2}}\) satisfies the assumptions of Theorem~2 and is
steerable from Alice to Bob.

\paragraph{The state lies outside the Theorem~1 class.}
Every state satisfying the structural assumptions of Theorem~1 has
rank at most three, as shown in
Eq.~\eqref{eq:link_rank_bound}. The rank bound is a property of the
state itself: it follows from the representation
\(\rho_{AB}=V\Gamma V^\dagger\) with \(\operatorname{rank}\Gamma\leq3\),
and is invariant under local basis changes on Alice and Bob. Since
\(\rho_{\mathrm{T2}}\) has rank five, it cannot satisfy the
structural assumptions of Theorem~1. This example therefore shows
that Theorem~2 applies to steerable states outside the state class
characterized by Theorem~1.

\subsection{Theorem-certified data and dataset augmentation}
\label{app:theorem_state_families}

We construct two sets of theorem-certified steerable states for
independent evaluation and, in a separate experiment, training-set
augmentation.

For Theorem~1, we sample states in the orthogonal form of
Eq.~\eqref{eq:link_explicit_matrix},
\begin{equation}
	\widetilde{\rho}_{AB}
	=
	\begin{pmatrix}
		\tilde{\mu}_1 & & & & a_{12} & & & & b_{13}\\
		& 0 & & & & & & & \\
		& & 0 & & & & & & \\
		& & & 0 & & & & & \\
		\overline{a_{12}} & & & & \tilde{\mu}_2 & & & & c_{23}\\
		& & & & & 0 & & & \\
		& & & & & & 0 & & \\
		& & & & & & & 0 & \\
		\overline{b_{13}} & & & & \overline{c_{23}} & & & & \tilde{\mu}_3
	\end{pmatrix},
	\label{eq:theorem1_sample_matrix}
\end{equation}
where blank entries are zero. We draw a complex Ginibre matrix
\(G\in\mathbb{C}^{3\times3}\) and set
\begin{equation}
	\Gamma
	=
	\frac{GG^\dagger}{\operatorname{tr}(GG^\dagger)}.
	\label{eq:theorem1_sample_gamma}
\end{equation}
The matrix \(\Gamma\) is positive semidefinite with unit trace by
construction. We read off
\begin{equation}
	\widetilde{\rho}_{AB}
	=
	\begin{pmatrix}
		\Gamma_{11} & & & & \Gamma_{01} & & & & \Gamma_{02}\\
		& 0 & & & & & & & \\
		& & 0 & & & & & & \\
		& & & 0 & & & & & \\
		\overline{\Gamma_{01}} & & & & \Gamma_{22} & & & & \Gamma_{12}\\
		& & & & & 0 & & & \\
		& & & & & & 0 & & \\
		& & & & & & & 0 & \\
		\overline{\Gamma_{02}} & & & & \overline{\Gamma_{12}} & & & & \Gamma_{33}
	\end{pmatrix}.
\end{equation}
which is Eq.~\eqref{eq:theorem1_sample_matrix} with
\(\tilde{\mu}_i=\Gamma_{ii}\), \(a_{12}=\Gamma_{01}\),
\(b_{13}=\Gamma_{02}\), and \(c_{23}=\Gamma_{12}\).
This guarantees that
\(\widetilde{\rho}_{AB}\) is positive and normalized. We retain
only samples with at least one nonzero coherence among
\(a_{12},b_{13},c_{23}\); by Theorem~1, these states are steerable.
This yields \(1000\) steerable \(\widetilde{\rho}_{AB}\).  Let \(\mathcal{S}_{\mathrm{T}1}\) denote the steerable samples obtained above.

For Theorem~2, we sample
\begin{equation}
	\rho_{\mathrm{T2}}
	=
	p_1|\psi(\theta,\phi)\rangle\langle\psi(\theta,\phi)|
	+p_2|10\rangle\langle10|
	+p_3|12\rangle\langle12|
	+p_4|20\rangle\langle20|
	+p_5|21\rangle\langle21|,
	\label{eq:theorem2_sample_state}
\end{equation}
where
\begin{equation}
	|\psi(\theta,\phi)\rangle
	=
	\sin\theta\cos\phi\,|00\rangle
	+\sin\theta\sin\phi\,|11\rangle
	+\cos\theta\,|22\rangle.
	\label{eq:theorem2_sample_pure_state}
\end{equation}
We draw
\((p_1,\ldots,p_5)\sim\operatorname{Dirichlet}(1,1,1,1,1)\)
and \(\theta,\phi\sim\operatorname{Uniform}(0,\pi/2)\)
independently. The sampled states satisfy the assumptions of
Theorem~2 and are therefore steerable. This yields \(1000\) steerable \(\rho_{\mathrm{T2}}\). Let \(\mathcal{S}_{\mathrm{T}2}\) denote the steerable samples obtained above.

Let~\(\mathcal{S}_{\mathrm{T}i}\)
denote the two labeled steerable state sets. To keep the classes balanced, for each \(i\), we independently generate \(1000\) separable states \(\mathcal{U}_i\), labeled \(+1\), and define the balanced sets
\begin{equation}
	\mathcal{D}_{\mathrm{aug}i}
	=
	\mathcal{S}_{\mathrm{T}i}\cup\mathcal{U}_i,
	\qquad i\in\{1,2\}.
	\label{eq:theorem_certified_test_sets}
\end{equation}
For \(\mathcal{D}_{\mathrm{cert}}\), we define
\begin{equation}
	\mathcal{D}_{\mathrm{cert}}
	=
	\mathcal{D}_{\mathrm{acc}}\cup\mathcal{D}_{\mathrm{aug}1}\cup\mathcal{D}_{\mathrm{aug}2}.
	\label{eq:theory_augmented_dataset}
\end{equation}

The material in S1--S5 is required by the main text: it defines the CEQRP features (S1), reports the classification
benchmarks (S2), proves Theorems~1 and~2 (S3),
establishes the relation between the two theorems (S4), and
specifies the theorem-certified datasets and their augmentation
(S5). Appendices S6 and S7 provide additional experimental results that complement the main text: S6 reports the performance of the augmented classifiers on measurement-number-independent datasets, and S7 examines the finite-measurement
steering boundaries for Werner and isotropic states. S8 collects the SDP verification details for the partially entangled states.

\subsection{Results on measurement-number-independent datasets}
\label{app:measurement_independent_results}

This section reports CEQRP classification results in two settings. Section~\ref{app:additional_experimental_acc} trains classifiers on the reference dataset \(\mathcal{D}_{\mathrm{acc}}\) alone and
evaluates them on independently certified states.
Section~\ref{app:additional_experimental_results-1} trains
classifiers on the theory-augmented dataset
\(\mathcal{D}_{\mathrm{cert}}\) and evaluates them on held-out Werner and PE states. The two settings together show that theory-certified augmentation
improves classification on the held-out Werner and PE families, relative to training on \(\mathcal{D}_{\mathrm{acc}}\) alone. The
augmented classifiers are therefore used in the main text to screen PE states for candidate steerability, with the candidates verified independently by SDP.

\subsubsection{Classifiers trained on \(\mathcal{D}_{\mathrm{acc}}\): generalization to theorem-certified states}
\label{app:additional_experimental_acc}

The reference training set is the accurately labeled dataset
\(\mathcal{D}_{\mathrm{acc}}\) of Ref.~\cite{h7qp-63dg}. Its
steerable class comprises steerable Werner and isotropic states, entangled pure states, and random states certified as steerable by SDP for \(m=3,\ldots,7\) measurement settings; its unsteerable class comprises separable states and unsteerable Werner and isotropic states. Steerable and unsteerable states carry labels
\(-1\) and \(+1\), respectively.

The three test sets \(\mathcal{D}_{\mathrm{aug1}}\),
\(\mathcal{D}_{\mathrm{aug2}}\), and \(\mathcal{D}_{\mathrm{PE}}\)
are those defined in Eqs.~\eqref{eq:theorem_certified_test_sets} and the PE paragraph of ~\ref{app:theorem_state_families}.

We evaluate generalization using accuracy (ACC), steerable-state recall (SR), precision (P), and the \(F_1\) score. The results for \(F_1\)-CEQRP-AdaBoost and \(F_2\)-CEQRP-MLP are reported in
Tables~\ref{tab:theory_certified_results_2}
and~\ref{tab:theory_certified_results_3}, respectively. These are the classifiers that performed best on the three independent test sets; they need not coincide with the best classifiers in~\ref{app:ceqrp_full_readout_results}, where the evaluation target was Werner states and isotropic states rather than theorem-certified states.

\begin{table}[htbp]
	\centering
	\begin{minipage}[t]{0.5\textwidth}
		\centering
		\footnotesize
		\caption{Performance of the \(F_1\)-CEQRP-AdaBoost classifier on the test sets.}
		\label{tab:theory_certified_results_2}
		\setlength{\tabcolsep}{2pt}
		\begin{tabular}{lcccc}
			\toprule
			Test set & ACC & SR & P & F1 \\
			\midrule
			\(\mathcal{D}_{\mathrm{aug}1}\) & 97.25\% & 100.00\% & 94.78\% & 97.32\% \\
			\(\mathcal{D}_{\mathrm{aug}2}\) & 80.06\% & 97.73\% & 72.12\% & 83.06\% \\
			\(\mathcal{D}_{\mathrm{PE}}\)     & 99.60\% & 100.00\% & 99.21\% & 99.60\% \\
			\bottomrule
		\end{tabular}
	\end{minipage}\hfill
	\begin{minipage}[t]{0.48\textwidth}
		\centering
		\footnotesize
		\caption{Performance of the \(F_2\)-CEQRP-MLP classifier on the test sets.}
		\label{tab:theory_certified_results_3}
		\setlength{\tabcolsep}{2pt}
		\begin{tabular}{lcccc}
			\toprule
			Test set & ACC & SR & P & F1 \\
			\midrule
			\(\mathcal{D}_{\mathrm{aug}1}\) & 99.95\% & 99.90\% & 100.00\% & 99.95\% \\
			\(\mathcal{D}_{\mathrm{aug}2}\) & 81.55\% & 99.75\% & 73.13\% & 84.39\% \\
			\(\mathcal{D}_{\mathrm{PE}}\)     & 94.50\% & 99.60\% & 90.38\% & 94.77\% \\
			\bottomrule
		\end{tabular}
	\end{minipage}
\end{table}

\subsubsection{Classifiers trained on \(\mathcal{D}_{\mathrm{cert}}\): effect of theorem-certified augmentation}
\label{app:additional_experimental_results-1}

We train classifiers on the augmented dataset
\(\mathcal{D}_{\mathrm{cert}}\)
defined in Eq.~\eqref{eq:theory_augmented_dataset}.

The PE test set \(\mathcal{D}_{\mathrm{PE}}\) is held out
throughout, and we additionally hold out a balanced Werner test set \(\mathrm{WernerTest}\), which is excluded from all training and model-selection steps. The \(F_i\)-CEQRP-ML classifiers trained on
\(\mathcal{D}_{\mathrm{cert}}\) are evaluated on
\(\mathrm{WernerTest}\) and on \(\mathcal{D}_{\mathrm{PE}}\).
Tables~\ref{tab33-2} and~\ref{tab33-3} report the results for \(F_1\)-CEQRP-ML and \(F_2\)-CEQRP-ML, respectively.

\begin{table}[!htbp]
	\centering
	\begin{minipage}[t]{0.48\textwidth}
		\centering
		\footnotesize
		\caption{Classification results for the \(F_1\)-CEQRP-ML classifiers trained on \(\mathcal{D}_{\mathrm{cert}}\).}
		\label{tab33-2}
		\setlength{\tabcolsep}{1pt}
		\begin{tabular}{lcccccc}
			\toprule
			Model & LR & SVM & MLP & GB & AdaBoost & EM \\
			\midrule
			CV             & 0.9983 & 0.9994 & 0.9975 & 0.9996 & 0.9995 & 0.9997 \\
			Validation     & 0.9965 & 0.9982 & 0.9995 & 0.9994 & 0.9997 & 0.9996 \\
			WernerTest    & 0.8755 & 0.9665 & 0.9858 & 0.9570 & 0.8155 & 0.8170 \\
			PETest  & 0.9585 & 0.9498 & 0.9273 & 0.9645 & 0.9371 & 0.9283 \\
			\bottomrule
		\end{tabular}
	\end{minipage}\hfill
	\begin{minipage}[t]{0.48\textwidth}
		\centering
		\footnotesize
		\caption{Classification results for the \(F_2\)-CEQRP-ML classifiers trained on \(\mathcal{D}_{\mathrm{cert}}\).}
		\label{tab33-3}
		\setlength{\tabcolsep}{1pt}
		\begin{tabular}{lcccccc}
			\toprule
			Model & LR & SVM & MLP & GB & AdaBoost & EM \\
			\midrule
			CV             & 0.9924 & 0.9935 & 0.9939 & 0.9879 & 0.9899 & 0.9976 \\
			Validation     & 0.9973 & 0.9985 & 0.9900 & 0.9894 & 0.9957 & 0.9950 \\
			WernerTest    & 0.9695 & 0.9865 & 0.9735 & 0.9748 & 0.8748 & 0.9748 \\
			PETest  & 0.9973 & 0.9843 & 0.9887 & 0.8490 & 0.7990 & 0.9347 \\
			\bottomrule
		\end{tabular}
	\end{minipage}
\end{table}

\subsection{Finite-Measurement Steering Boundaries for Werner states and Isotropic States}

As a complement to the measurement-setting hierarchy of PE-state steering examined in the main text, we perform the same across-\(m\) comparison for the Werner and isotropic reference families: for each \(m=3,\ldots,7\), the CEQRP-predicted boundaries are compared with the corresponding finite-measurement SDP thresholds. The results are shown in Figs.~\ref{fig:werner_m_boundary_SM_1} and \ref{fig:isotropic_m_boundary_SM_2}, respectively.

\begin{center}
	\begin{minipage}[htbp]{0.48\textwidth}
		\centering
		\includegraphics[width=\linewidth,height=0.32\textheight,keepaspectratio]{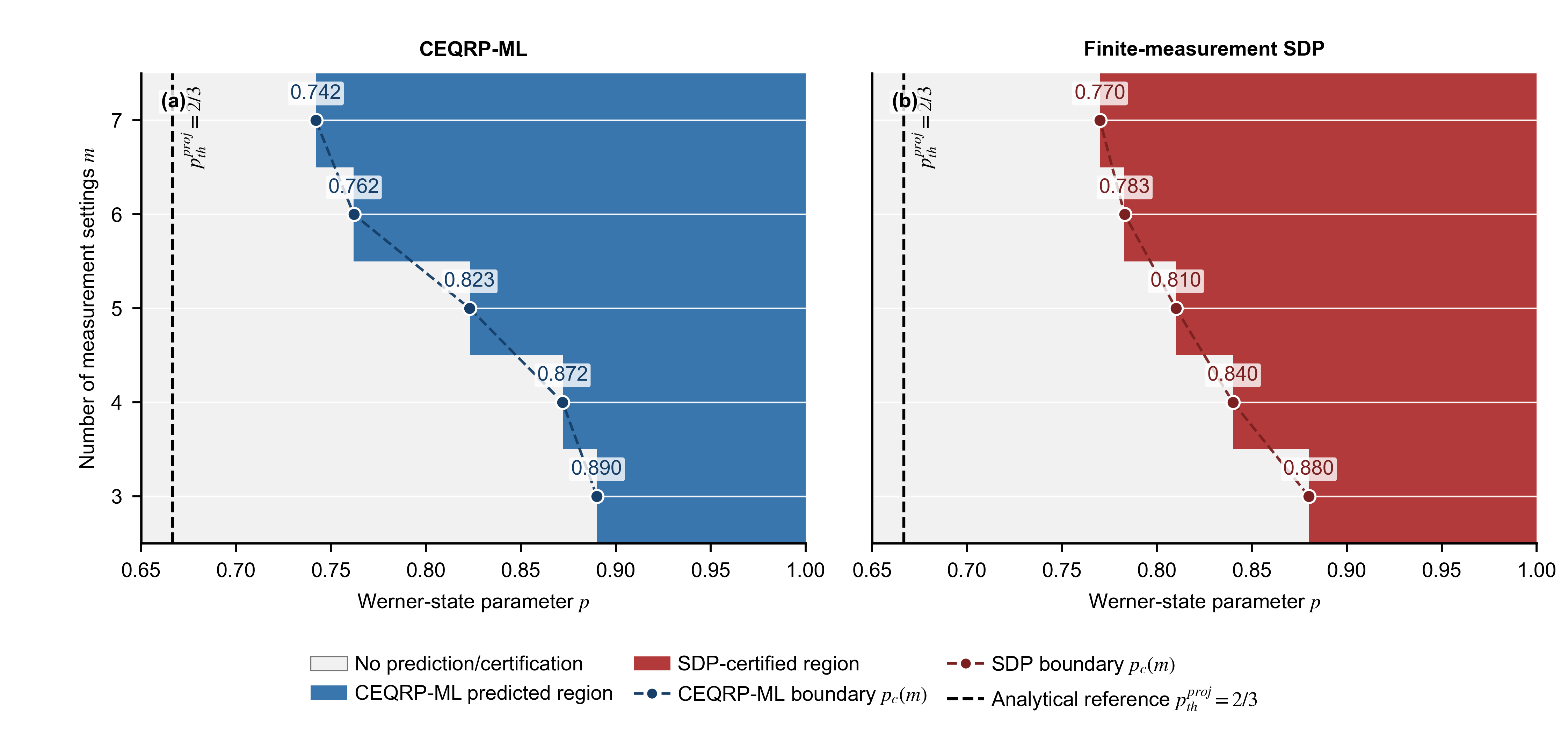}
		\captionof{figure}{Finite-measurement steering maps for Werner states.
			The horizontal axis denotes the state parameter \(p\), and the
			vertical axis denotes the number of measurement settings
			\(m\in\{3,4,5,6,7\}\). Panel~(a) shows the steering candidates
			predicted by the CEQRP-ML models, while panel~(b) shows the
			steering region certified by finite-measurement semidefinite
			programming (SDP). The markers indicate the critical boundary
			\(p_{\mathrm{c}}(m)\) for each fixed \(m\).}
		\label{fig:werner_m_boundary_SM_1}
	\end{minipage}\hfill
	\begin{minipage}[t]{0.48\textwidth}
		\centering
		\includegraphics[width=\linewidth,height=0.3\textheight,keepaspectratio]{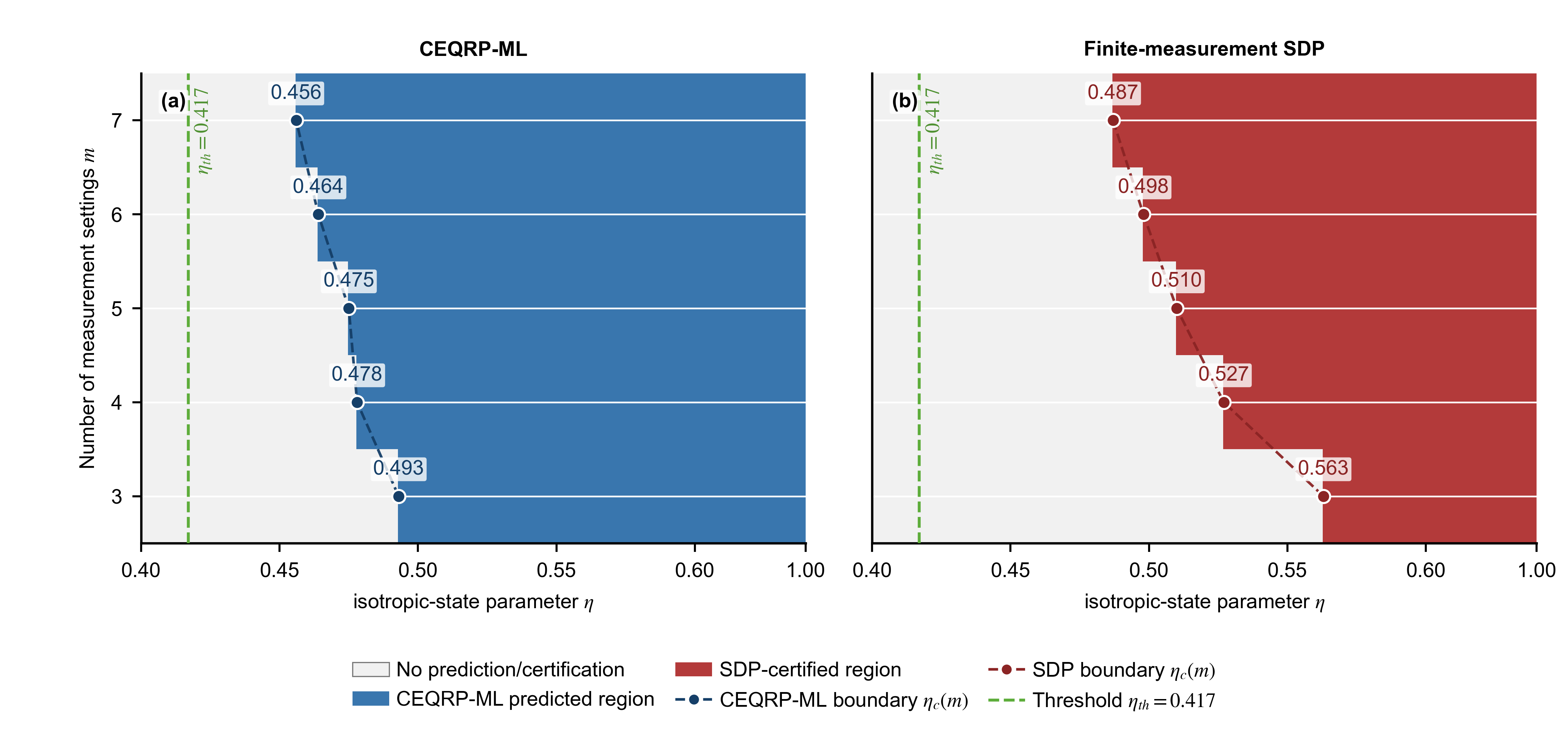}
		\captionof{figure}{Finite-measurement steering maps for isotropic states.
			The horizontal axis denotes the state parameter \(\eta\), and
			the vertical axis denotes the number of measurement settings
			\(m\in\{3,4,5,6,7\}\). Panel~(a) shows the steering candidates
			predicted by the CEQRP-ML models, while panel~(b) shows the
			steering region certified by finite-measurement SDP. The markers
			indicate the critical boundary \(\eta_{\mathrm{c}}(m)\) for each
			fixed \(m\).}
		\label{fig:isotropic_m_boundary_SM_2}
	\end{minipage}
\end{center}

\subsection{SDP Verification of Steering for Partially Entangled States: Feasible Measurement Directions}
\label{measurement}

For the PE state at the specified parameter values, we give the measurement directions used to certify steering with the SDP method at \(\theta=\pi/4\), \(\phi=1.5101\), and \(p=0.45\). The remaining validation and measurement results are available in the GitHub repository cited in Ref.~\cite{ref45}.

\begin{table}[H]
	\centering
	\caption{Feasible measurement directions for SDP verification of
		PE-state steering, with \(\theta=\pi/4\), \(\phi=1.5101\), and
		\(p=0.45\).}
	\label{tab:measurement_directions}
	\footnotesize
	\setlength{\tabcolsep}{1pt}
	\begin{tabular}{clccc}
		\toprule
		\# & \(\mathbf{n}\) & \(\|\mathbf{n}\|\) & \(\theta\) (rad) & \(\phi\) (rad) \\
		\midrule
		1 & \((0.0017032835,\ 0.2998181112,\ -0.9539948632)\) & 1.00000000 & 2.83708559 & 1.56511533 \\
		2 & \((0.2393197383,\ 0.7801306155,\ -0.5780331181)\) & 1.00000000 & 2.18711259 & 1.27314124 \\
		3 & \((-0.0457672755,\ -0.9385867021,\ 0.3419946770)\) & 1.00000000 & 1.22175758 & 4.66366567 \\
		4 & \((0.6083754950,\ 0.3564404343,\ 0.7091046988)\) & 1.00000000 & 0.78256867 & 0.52997905 \\
		5 & \((0.7426075032,\ -0.1132575675,\ -0.6600809190)\) & 1.00000000 & 2.29172280 & 6.13183821 \\
		6 & \((-0.0943652618,\ 0.0257080332,\ -0.9952056543)\) & 1.00000000 & 3.04363163 & 2.87561628 \\
		7 & \((-0.2531736545,\ -0.3569812728,\ -0.8991481922)\) & 1.00000000 & 2.68861559 & 4.09551044 \\
		\bottomrule
	\end{tabular}
\end{table}

\end{document}